\documentclass[prd,letterpaper,twocolumn,superscriptaddress,preprintnumbers,nofootinbib]{revtex4}

\pdfoutput=1
\usepackage{slashed}
\usepackage{amsmath,amssymb}
\usepackage{graphicx}
\usepackage{units}
\usepackage{bbold}
\usepackage{xcolor}
\usepackage{dsfont}
\usepackage{enumerate}
\usepackage{xcolor}

\definecolor{naviblue}{rgb}{0.0, 0.0, 0.5}

\usepackage[colorlinks  = true,
            linkcolor   = naviblue,
            urlcolor    = naviblue,
            citecolor   = naviblue,
            anchorcolor = naviblue]{hyperref}

\usepackage{comment}
\usepackage[normalem]{ulem}   
\usepackage{multirow}

\newcommand{\beq}{\begin{equation}}
\newcommand{\eeq}{\end{equation}}
\newcommand{\bea}{\begin{eqnarray}}
\newcommand{\ena}{\end{eqnarray}}

\def \epsilon {\varepsilon} 

\def \vec#1{{\boldsymbol{#1}}}

\newcommand{\ie}{\emph{i.e.}}
\newcommand{\eg}{\emph{e.g.}}

\newcommand{\miss}{\mathrm{miss}}
\newcommand{\calo}{\mathrm{calo}}
\newcommand{\trans}{\mathrm{T}}
\newcommand{\obs}{\mathrm{obs}}
\newcommand{\SM}{\mathrm{SM}}
\newcommand{\LLP}{\mathrm{LLP}}

\newcommand{\mDV}{m_\mathrm{DV}}
\newcommand{\nT}{n_{\rm trk}}
\newcommand{\met}{E_\trans^\miss}
\newcommand{\trk}{{\rm trk}}
\newcommand{\dEdx}{\mathrm{d}E/\mathrm{d}x}

\begin{document}

\title{Probing conversion-driven freeze-out at the LHC}

\author{Jan Heisig}
\email[E-mail: ]{heisig@physik.rwth-aachen.de}
\affiliation{Institute for Theoretical Particle Physics and Cosmology (TTK), RWTH Aachen University, D-52056 Aachen, Germany}
\affiliation{Department of Physics, University of Virginia,
Charlottesville, Virginia 22904-4714, USA}

\author{Andre Lessa}
\email[E-mail: ]{andre.lessa@ufabc.edu.br}

\affiliation{Centro de Ciências Naturais e Humanas, Universidade Federal do ABC,
Santo André, 09210-580 SP, Brazil}

\author{Lucas Magno D. Ramos}
\email[E-mail: ]{lucas.magno.ramos@usp.br}
\affiliation{Instituto de Física, Universidade de São Paulo, C.P. 66.318, 05315-970 São Paulo, Brazil}

\preprint{TTK-23-13}

\begin{abstract}
Conversion-driven freeze-out is an appealing mechanism to explain the observed relic density while naturally accommodating the null-results from direct and indirect detection due to a very weak dark matter coupling. Interestingly, the scenario predicts long-lived particles decaying into dark matter with lifetimes favorably coinciding with the range that can be resolved at the LHC\@. However, the small mass splitting between the long-lived particle and dark matter renders the visible decay products soft, challenging current search strategies. We consider four different classes of searches covering the entire range of lifetimes: heavy stable charge particles, disappearing tracks, displaced vertices, and missing energy searches. We discuss the applicability of these searches to conversion-driven freeze-out and derive current constraints highlighting their complementarity. For the displaced vertices search, we demonstrate how a slight modification of the current analysis significantly improves its sensitivity to the scenario.
\end{abstract}

\maketitle

\section{Introduction}
\label{sec:intro}

Conversion-driven freeze-out 
(CDFO)  or coscattering~\cite{Garny:2017rxs,DAgnolo:2017dbv}
constitutes an appealing  mechanism to explain the measured dark matter (DM) density~\cite{Planck:2018vyg} going beyond the widely studied paradigm of weakly interacting particles (WIMPs). The considerably smaller couplings required by this scenario render it consistent with the null-results of WIMP searches. They also support out-of-equilibrium conditions in the early Universe offering a compelling link to baryogenesis~\cite{Heisig:2024mwr}. Interestingly, the mechanism predicts long-lived particles (LLPs) at the LHC, typically with decay lengths of the order of $(10^{-3}\!-\!1)$\,m.
Unlike other feebly coupled DM scenarios yielding LLPs, the conversion-driven freeze-out mechanism requires DM masses below a few TeV, potentially exposing its entire parameter space to current and future collider searches. 

A simple realization of the mechanism can be found in so-called $t$-channel mediator DM models supplementing the standard model (SM) with a gauge singlet DM particle $X$ and a coannihilating partner $Y$ that mediates the DM interactions with the SM via a Yukawa coupling involving $X,Y$ and a SM fermion, \eg~a SM quark as considered here. Intriguingly, within this class of models, the WIMP region is almost entirely excluded by the complementary constraints from the relic density and indirect and direct DM detection searches~\cite{Arina:2023msd} rending the conversion-driven freeze-out region to be the prime target for investigations within those models at the LHC\@.

For the considered case of a quark-philic model, the mediator-pair production cross section at the LHC is sizeable. Its non-prompt decay potentially gives rise to a prominent LLP signature  including the (anomalous) track of the $Y$ as well as the displaced quark and the missing energy from its decay, $Y\to X q$. 
However, while the predicted range of lifetimes coincides very favorably with the range of displacements resolvable by LHC detectors the relatively small mass splittings between $Y$ and DM particle $X$ required by the mechanism  -- typically a few tens of GeV -- can pose severe  challenges for probing the scenario with current search strategies, specifically for intermediate to small lifetimes. The quark coming from the decay of the produced mediator is relatively soft, roughly of the order of the mass splitting. This signature of soft displaced jets constitutes a blind spot in the current search program.

In this paper, we discuss the sensitivity of several search strategies to this scenario covering the entire range of possible lifetimes:\footnote{ 
This list only contains references to the searches used in this study. In the given categories, other full-luminosity LLP searches can be found in Ref.~\cite{CMS:2024trg,CMS:2023mny,ATLAS:2022rme,ATLAS-EXOT-2019-23,CMS:2021tkn,CMS:2019ybf}.
We would also like to mention the searches for delayed jets~\cite{CMS:2019qjk}  and mildly-displaced low-momentum tracks~\cite{ATLAS:2024umc}, which, however, are not considered here, as we do not expect them to significantly change our results.}
\begin{itemize}
\item Searches for heavy stable charged particles through anomalous ionization and time-of-flight measurements are fairly inclusive relying only on the track of a slow-moving (heavy) charged particle~\cite{ATLAS-SUSY-2018-42}. Current searches are, however, only sensitive to decay lengths larger than around a meter.

\item Searches for disappearing tracks, which target charged tracks without hits in the tracker outer layers~\cite{CMS-EXO-19-010,CMS-EXO-16-044}. These searches have focused on Wino/Higgsino-like DM, where  an ultra-soft pion is radiated in the decay of the chargino. These searches are typically most sensitive for decay lengths of around a few tens of centimeters.

\item Searches for displaced jets have been performed with or without missing energy~\cite{ATLAS-SUSY-2016-08,ATLAS-SUSY-2018-13}. Potentially they can provide promising sensitivity in the range 4~mm \,$\lesssim c \tau \lesssim$\,30~cm.

\item Finally, missing energy searches~\cite{CMS-EXO-20-004} may compete for very small decay lengths, \ie~$ c \tau \lesssim\,$mm. 
Their sensitivity may, however, be extended towards larger lifetimes if the displaced objects are too soft and are considered as pile up.
\end{itemize}
We critically assess the capability of the current search program to probe CDFO for all four classes of searches above and suggest ways to close the gaps and improve on the current searches to better target the CDFO scenario.
Earlier efforts addressing this task can be found  in~\cite{Brooijmans:2020yij}. 

The remainder of the paper is structured as follows. In Sec.~\ref{sec:model}, we introduce the model considered in this work and discuss its cosmological constraints. Sections~\ref{sec:lhc} and \ref{sec:currconstr} discuss the model signatures and derive the current constraints from searches at the LHC, respectively. Finally, we propose a modified displaced jet search and demonstrate the significant improvement of sensitivity towards small to intermediate lifetimes in Sec.~\ref{sec:propse} before concluding in Sec.~\ref{sec:concl}. Appendix~\ref{app:cfl}
lists the cut flows of the considered analyses and Appendix~\ref{app:lhcHL} presents estimates for the high luminosity LHC.

The recasting code, input files and additional information required for reproducing the results presented here are available in the \href{https://doi.org/10.5281/zenodo.11536735}{Zenodo}~\cite{zenodoRepo} and \href{https://github.com/andlessa/LLP-CDFO}{GitHub}~\cite{githubRepo} repositories.

\section{Benchmark model and viable parameter space}
\label{sec:model}

We consider an extension of the SM by a singlet Majorana Fermion $X$ and a scalar mediator $Y$ both of which are odd under a new $Z_2$-symmetry. The mediator $Y$ has the same gauge quantum numbers as a right-handed SM quark, $q_R$, allowing for the Yukawa-type interaction
\begin{equation}
  {\cal L}_{\text{int}} = \lambda_X  Y  \bar q_R  X + \text{h.c.}\,.
\end{equation}
Additionally, the mediator interacts via the $SU(3)_c$ and $U(1)_Y$ SM gauge fields. The symmetries also allow for a Higgs portal interaction of $Y$, whose impact can, however, be neglected for our study assuming that the respective coupling is small compared with the strong coupling. We assume $m_Y>m_X$ rendering $X$ a stable DM candidate. For concreteness, we consider $q = b$, \ie~the mediator couples to the $b$ quark only. This model belongs to the class of $t$-channel mediator models (see \eg~\cite{Garny:2015wea}) which are being actively studied in the context of LHC DM searches~\cite{Arina:2023msd}.

Despite the model's simplicity, it provides a rich variety of mechanisms to explain the relic density ranging from canonical freeze-out to scenarios of non-thermalized DM\@. We focus on the regime of conversion-driven freeze-out requiring a very small DM coupling, $\lambda_X$, typically of the order of $10^{-6}$~\cite{Garny:2017rxs}. In this regime, the DM annihilation rate is insignificant over the entire DM freeze-out process such that the DM abundance is solely governed by conversion processes between $X$ and $Y$ which include decays and inverse decays $Y\leftrightarrow X q$ as well as scatterings off SM particles, \eg~$ Y q \leftrightarrow g X$. In fact, in conversion-driven freeze-out their rates are semi-efficient, such that conversions initiate the chemical decoupling of DM and hence determine the freeze-out abundance. This mechanism opens up a viable parameter space towards small mass splitting $\Delta m\equiv m_Y-m_X$, typically up to a few tens of GeV -- a region in which efficient conversions would lead to underabundant DM\@.
Interestingly, while avoiding constraints from direct and indirect detection, the required very small coupling is still large enough to thermalize DM, distinguishing this scenario from the regime of non-thermalized DM where the abundance may be explained by freeze-in~\cite{McDonald:2001vt,Hall:2009bx} or superWIMP production~\cite{Covi:1999ty,Feng:2003uy}.

\begin{figure}
    \centering
    \includegraphics[width=0.47\textwidth,trim={0.6cm 0.7cm 0.5cm 0.5cm},clip]{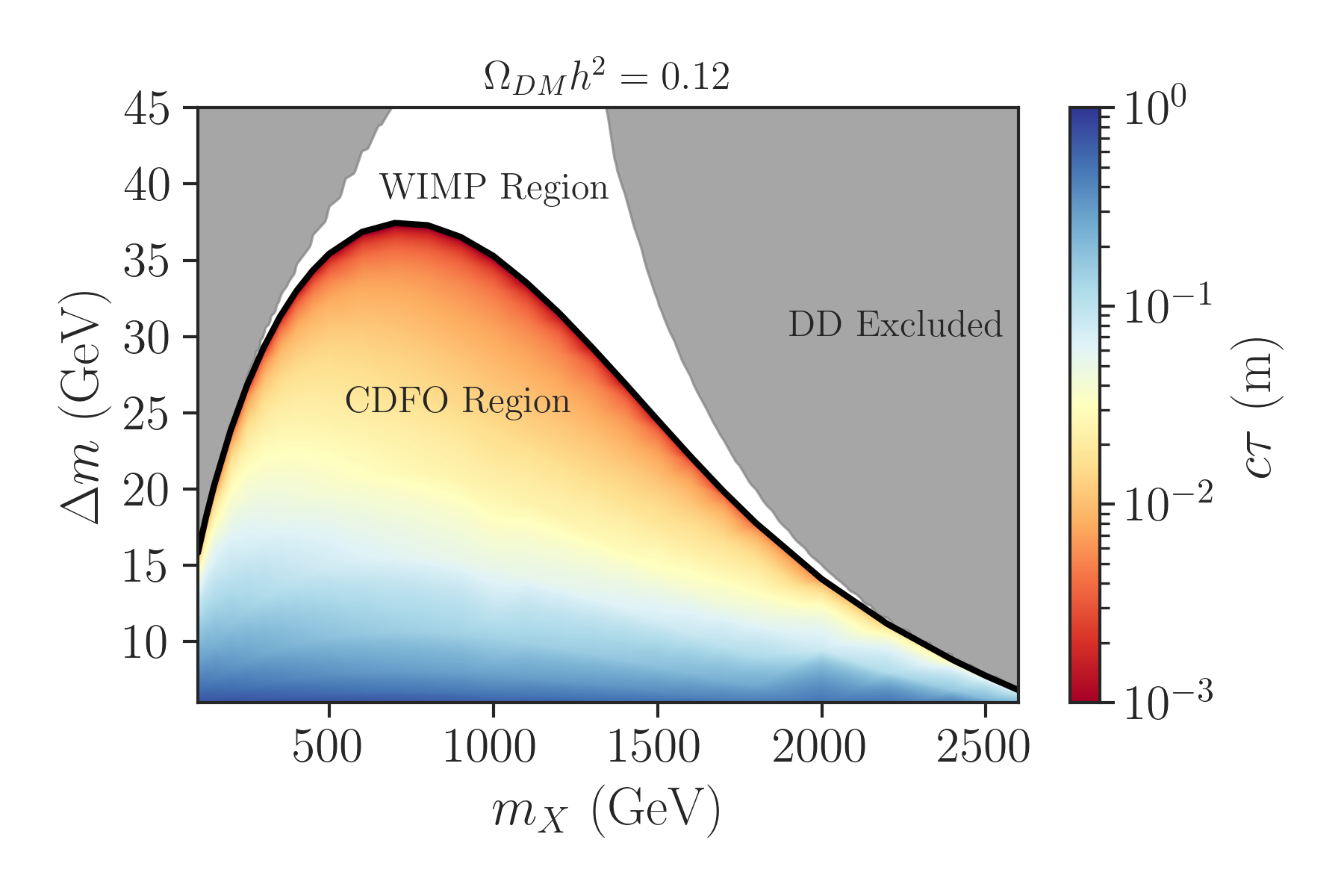} 
    \caption{Cosmologically viable parameter space, $\Omega_{DM} h^2 = 0.12$, in the WIMP region (above the black solid line) and conversion-driven freeze-out region (below). The color code denotes the $Y$ decay length in the latter region. The gray shaded area shows the 90\% CL direct detection exclusion from LZ~\cite{LZ:2022lsv}.}
    \label{fig:cosmoparam}
\end{figure}

The cosmologically viable parameter space, $\Omega h^2=0.12$~\cite{Planck:2018vyg} is shown in Fig.~\ref{fig:cosmoparam}.
This result was obtained using and extending the computation from Ref.~\cite{Garny:2021qsr}. In particular, we solved the
coupled set of Boltzmann equations for the abundances of $X$ and $Y$ taking into account Sommerfeld enhancement and bound state formation effects considering excitations up to a principal quantum number $n=15$ in the approximation of inefficient bound-to-bound transitions, see Ref.~\cite{Garny:2021qsr} for details.\footnote{While the inclusion of bound-to-bound transitions can have a large effect at small temperatures~\cite{Binder:2023ckj}, they have a minor effect on the parameter space of interest here, \ie~$m_X \lesssim1\,$TeV, as late annihilations are less important towards smaller DM masses.} Bound state effects are highly important for the freeze-out dynamics in this scenario, largely extending the viable parameter space, in particular, towards large DM masses.

The black solid line in Fig.~\ref{fig:cosmoparam} denotes the boundary between the WIMP region (above) and the conversion-driven freeze-out region (below). At the boundary the required coupling drops by many orders of magnitude, 
from the weak coupling regime with $\lambda_X \sim 0.1$ to the CDFO regime with $\lambda_X\sim 10^{-6}$.
The color code shows the resulting decay length of the mediator $Y$ in the latter region ranging from millimeters to around a meter. 
Note that in our analysis we restrict ourselves to the region $\Delta m \ge 6$\,GeV for which the 2-body decay of the mediator is open. For $\Delta m$ below the mass of the lightest $B$-meson, $m_B\simeq5.3\,$GeV, the mediator decay is four-body (or loop) suppressed rendering the mediator detector-stable.
In the WIMP region decays are prompt. The gray shaded regions denote 90\% CL direct detection exclusion region from LZ~\cite{LZ:2022lsv}, which are only relevant in the WIMP regime. We use \textsc{MadDM}~\cite{Ambrogi:2018jqj} for the computation of the direct detection cross section.

Interestingly, the viable CDFO parameter space is restricted to $m_Y\lesssim 4$\,TeV~\cite{Garny:2021qsr} (the maximum of which is reached in the region $\Delta m\to 0$ not displayed here) and hence defines a fully testable target for upcoming and future collider searches. 
This constitutes another difference to the case of non-thermalized DM where -- within the considered model -- the mediator can be as heavy as $10^9$\,GeV~\cite{Decant:2021mhj} leaving the bulge of the parameter space inaccessible for tests at laboratory experiments in the foreseeable future.

While focusing on the case $q=b$ here, we mention that the cases $q=u,d,s,c$ are expected to provide very similar results. 
The main difference concerns the position of the above-mentioned two-body decay threshold, which is at a smaller $\Delta m$ for the lighter quarks.
In addition, for the first-generation quarks, direct detection limits are even stronger than the ones shown in Fig.~\ref{fig:cosmoparam}, excluding almost the entire WIMP region~\cite{Arina:2023msd}. The case $q=t$ leads to a qualitatively different phenomenology, as $m_t>\Delta m$ in the entire conversion-driven freeze-out regime, see Ref.~\cite{Garny:2018icg}.

\section{Signal at the LHC}
\label{sec:lhc}

\begin{figure}
    \centering
    \includegraphics[width=0.47\textwidth,trim={0.25cm 0.33cm 0.0cm 0.33cm},clip]{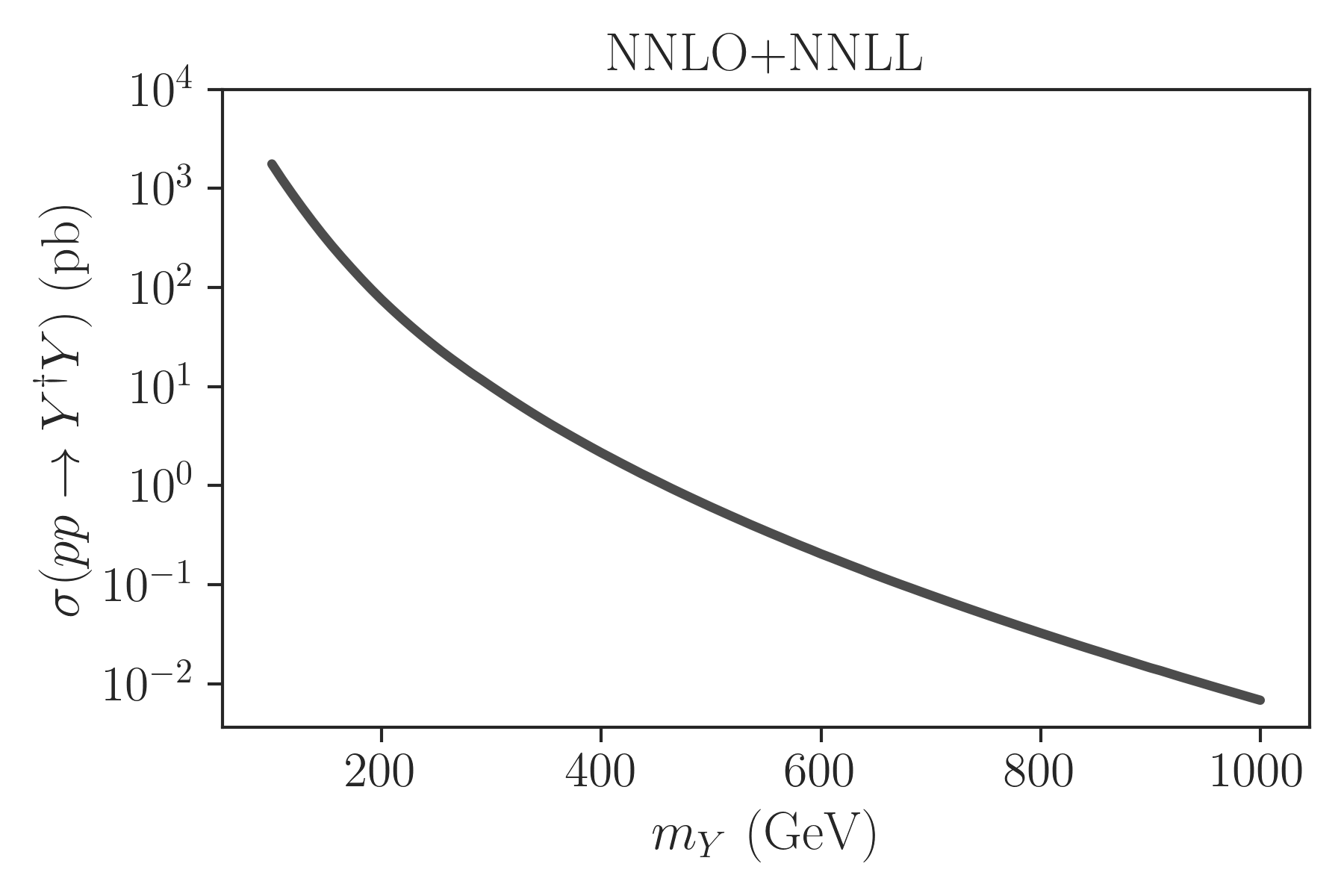}    
    \caption{Total production cross-section for $p p \to Y^\dagger Y$ at the LHC with a center of mass energy of 13 TeV. The cross-section is computed at NNLO$+$NNLL order using \textsc{NLLfast}~\cite{nnlfast}.}
    \label{fig:totalxsec}
\end{figure}

As the mediator $Y$ is color-charged, its pair production cross section at the LHC is sizeable even at $m_Y \sim$~TeV, as shown in Fig.~\ref{fig:totalxsec}. 
Before verifying the LHC constraints from current ATLAS and CMS searches, it is useful to identify the main features of the CDFO signal which can be relevant for these searches.
To compute the signal distributions, we generate Monte Carlo (MC) events at tree level for the process $p p \to Y^\dagger Y + 0,1$~jets, with $Y \to b X$. We use the MLM matching scheme with a matching scale set to $Q_{cut}= m_{Y}/5$. 
The events are generated using \textsc{MadGraph5\_aMC@NLO}~\cite{Alwall:2014hca,Frederix:2018nkq} for the parton level process with the PDF set \textsc{NNPDF23\_nlo\_as\_0130}, \textsc{ Pythia~8}~\cite{Bierlich:2022pfr} for showering and hadronization and a modified \textsc{Delphes}~\cite{Delphes} version for detector simulation and jet clustering.\footnote{The changes in \textsc{Delphes} were made to include additional information related to long-lived particles, which are relevant for computing the LHC constraints.} 
The total cross-section is normalized to its NNLO value, which was computed using \textsc{NLLfast}~\cite{nnlfast}.

\begin{figure*}
    \centering
    \includegraphics[width=0.48\textwidth,trim={0.25cm 0.2cm 0.0cm 0.4cm},clip]{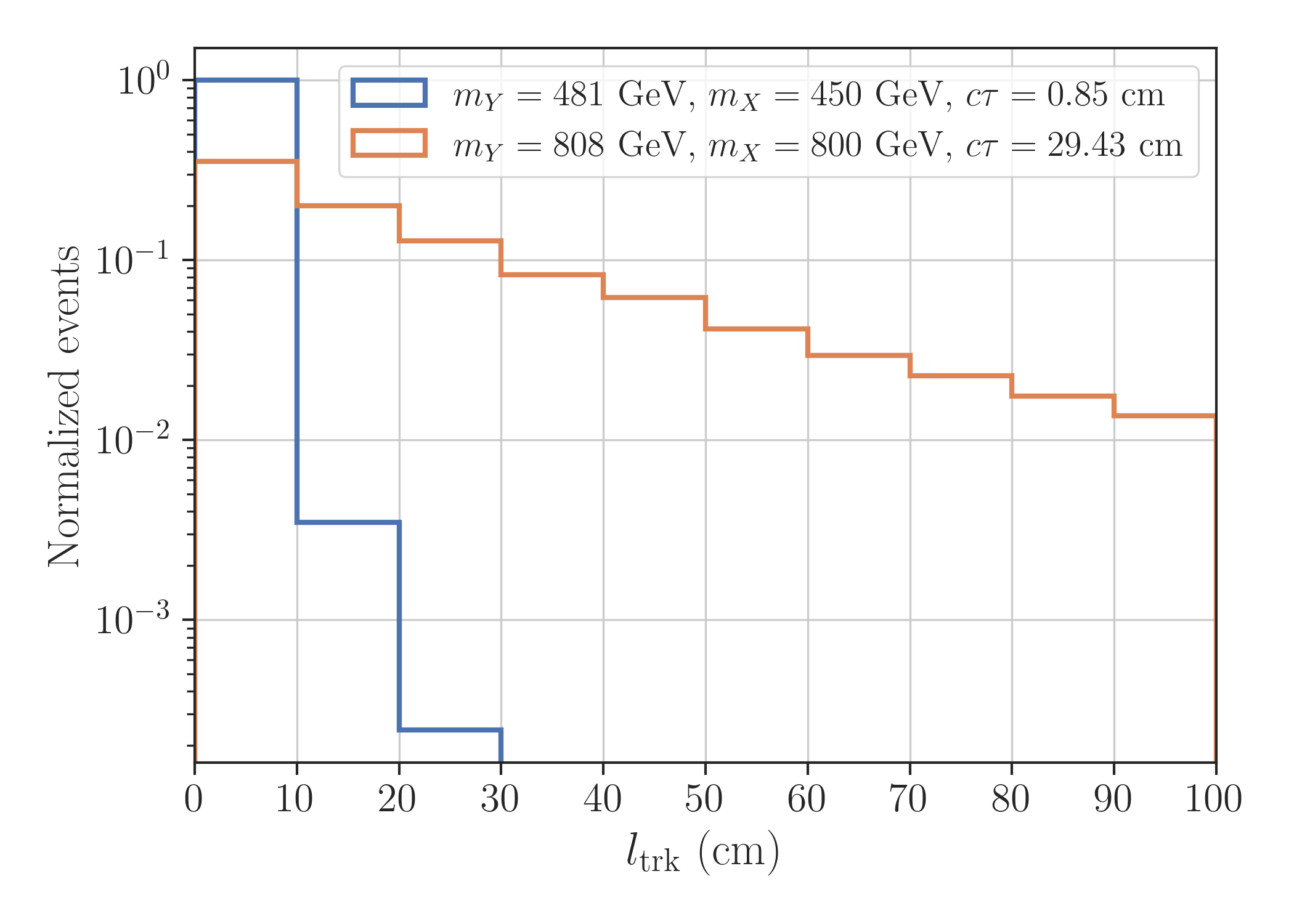}
    \includegraphics[width=0.48\textwidth,trim={0.45cm 0.2cm 0.0cm 0.4cm},clip]{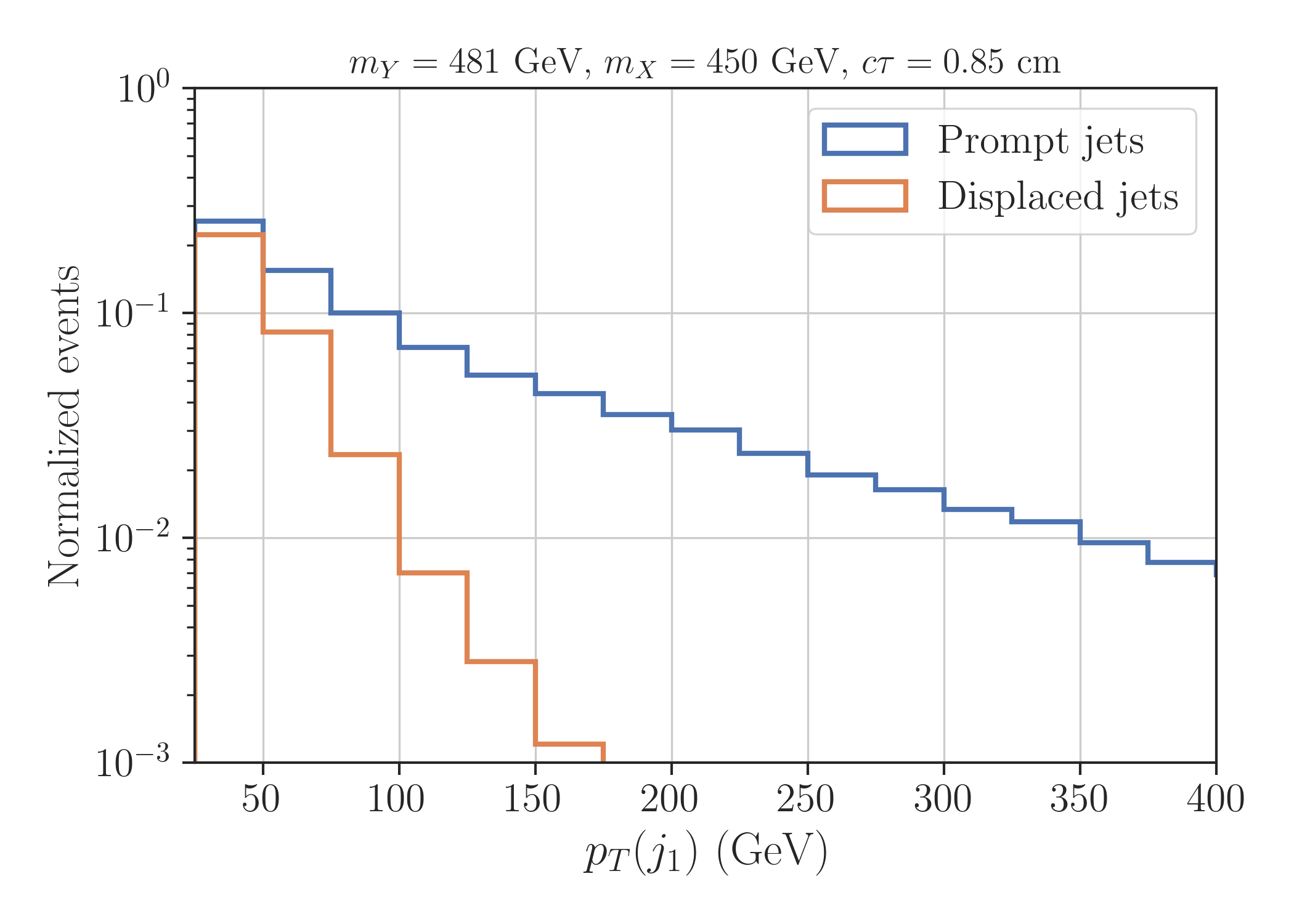}
    \includegraphics[width=0.48\textwidth,trim={0.45cm 0.6cm 0.0cm 0.2cm},clip]{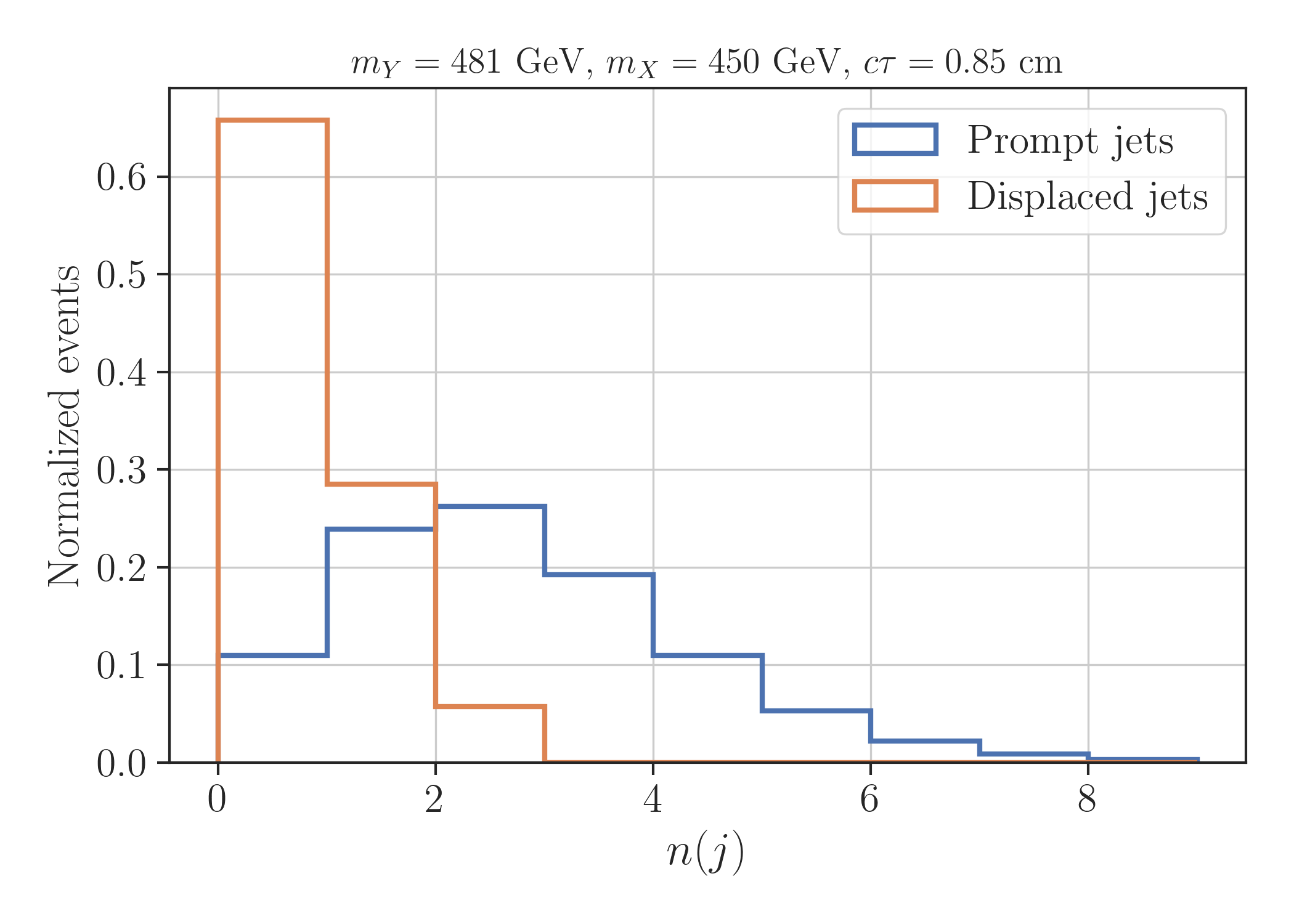}
    \includegraphics[width=0.48\textwidth,trim={0.45cm 0.6cm 0.0cm 0.2cm},clip]{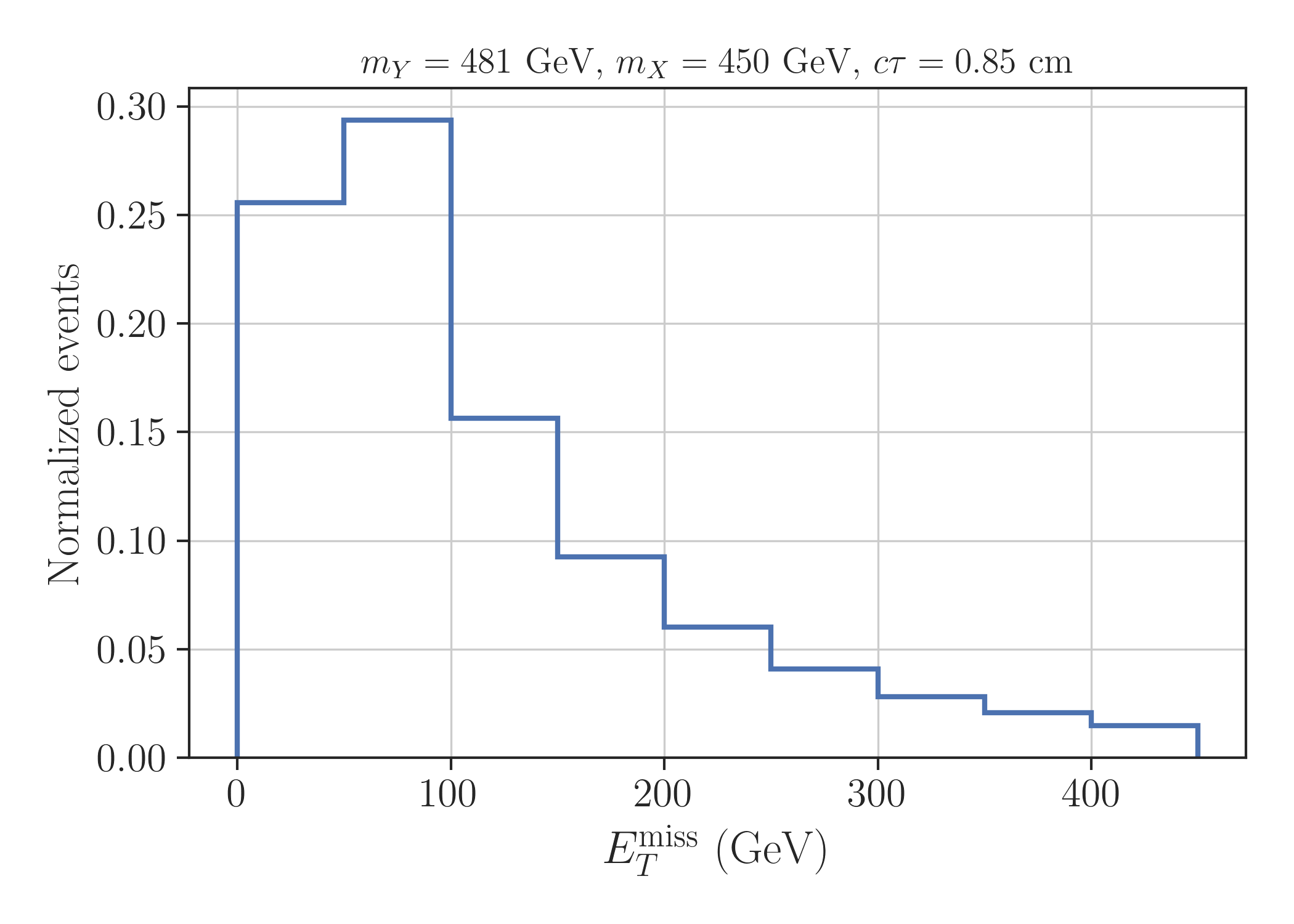}
    
    \caption{Kinematical distributions for the 13\,TeV LHC\@. 
    \textit{Upper left:} Decay length distribution (in the lab frame) for charged R-hadrons. The blue histogram shows the distribution for a benchmark with $c \tau \simeq 1$~cm, while the orange histogram has $c \tau \simeq 30$~cm.
    \textit{Upper right:} Jet $p_\trans$ distributions for prompt (blue) and displaced (orange) jets. The jets satisfy $p_\trans(j) > 25$~GeV and $|\eta(j)| < 5$.
    \textit{Lower left:} Number of jets distributions for prompt (blue) and displaced (orange) jets. The jets satisfy $p_\trans(j) > 25$~GeV and $|\eta(j)| < 5$.
    \textit{Lower right:} Missing energy distribution.
    }
    \label{fig:BMdistrib} 
\end{figure*}

As discussed in Sec.~\ref{sec:model}, the cosmologically viable parameter space within the CDFO predicts a macroscopic mediator decay length, ${\cal O}$(mm)\,$<c\tau < {\cal O}$(m). 
In this region, the pair-produced mediators form R-hadrons~\cite{Fairbairn:2006gg} from which around $44\%$ are electrically charged.\footnote{
The $Y$ hadronization and decay is computed using the default R-hadron implementation of \textsc{Pythia}~8.}
As seen in Fig.~\ref{fig:cosmoparam}, the mediator decay length rapidly decreases with $\Delta m$. The decay length distributions for two benchmarks with $\Delta m = 31$~GeV and $\Delta m = 8$~GeV are shown in Fig.~\ref{fig:BMdistrib} (upper left). For small $\Delta m$, 
a sizeable fraction of charged R-hadrons decay outside the tracker ($l_\trk \gtrsim 50$~cm),  thus leading to highly ionizing tracks.
The fraction of charged R-hadrons decaying within the tracker, but still traversing a sufficient number of inner layers, can lead to disappearing tracks since their decays mostly lead to missing energy.
For larger $\Delta m$ values, the $b$-jets from the R-hadron decays can carry more energy, leading to events with displaced $b$-jets and missing energy.
However, since $\Delta m < 40$~GeV is required by the CDFO mechanism, the displaced jets are relatively soft and pose a challenge to searches for this signature. 
Thus, to satisfy minimum trigger requirements, hard jets from initial state radiation (ISR) are usually needed.

Since the signal contains displaced jets from the R-hadron decays and prompt jets from ISR, it is relevant to distinguish between those. This is not always easy to achieve within the experiment, since soft and displaced tracks can be assumed to originate from pile up and removed from the event. 
In addition, the definition of displaced and prompt jets is analysis-dependent. 
Nonetheless, as a way to illustrate the signal features, we identify jets as displaced if they have
a small $\Delta R$ separation to one of the $b$ quarks from the $Y$ decay. The specific requirements for tagging a jet as displaced are:
\begin{align}
    p_\trans(j) > 20 \mathrm{ GeV}, \; \max_{i} \Delta R(j,b^{(i)}) < 0.3, \;
    R_{xy}(j) > 4 \mathrm{ mm},
\end{align}
where $b^{(i)}, i=1,2$ represents the $b$ quarks from $Y$ decays.
The transverse jet displacement, $R_{xy}(j)$, is then taken to be the same as the displacement of the closest $b$ quark. We point out that this prescription is similar to the one used for recasting the displaced jet search from Ref.~\cite{ATLAS-SUSY-2018-13}. The jets which do not satisfy the above criteria are then classified as {\it prompt}.

In Fig.~\ref{fig:BMdistrib} (upper right) we show the $p_\trans$ distributions for the displaced and prompt jets for a benchmark point with $m_{Y} = 481$~GeV, $\Delta m = 31$~GeV and a coupling $\lambda_X = 3.9 \times 10^{-7}$ required by $\Omega h^2 = 0.12$. The point's (proper) decay length is $c \tau = 0.85$~cm and its production cross section at the 13\,TeV LHC is $\sigma(Y^\dagger Y) = 0.76$~pb\@.
As expected, the displaced jets have a much softer distribution, due to the mass compression.
The number of displaced jets in each event is shown in Fig.~\ref{fig:BMdistrib} (lower left) and is significantly smaller than the number of prompt jets once we impose the minimal jet requirements: $p_\trans(j) > 25$~GeV and $|\eta(j)| < 5$. While the largest fraction of events does not contain any (hard) displaced jets, the number of prompt jets peaks at $n(j) = 3$.
Since $Y$ also decays to DM, we expect the signal to contain a sizeable fraction of missing energy.
Fig.~\ref{fig:BMdistrib} (lower right) shows the $\met$ distribution for the same benchmark. Despite the $Y$ mass being $481$~GeV, the missing energy spectrum peaks below 100~GeV, due to the small $\Delta m$, resulting in a softer distribution when compared to non-compressed models.

\begin{figure}
    \centering
    \includegraphics[width=0.48\textwidth,trim={0.45cm 0.6cm 0.4cm 0.4cm},clip]{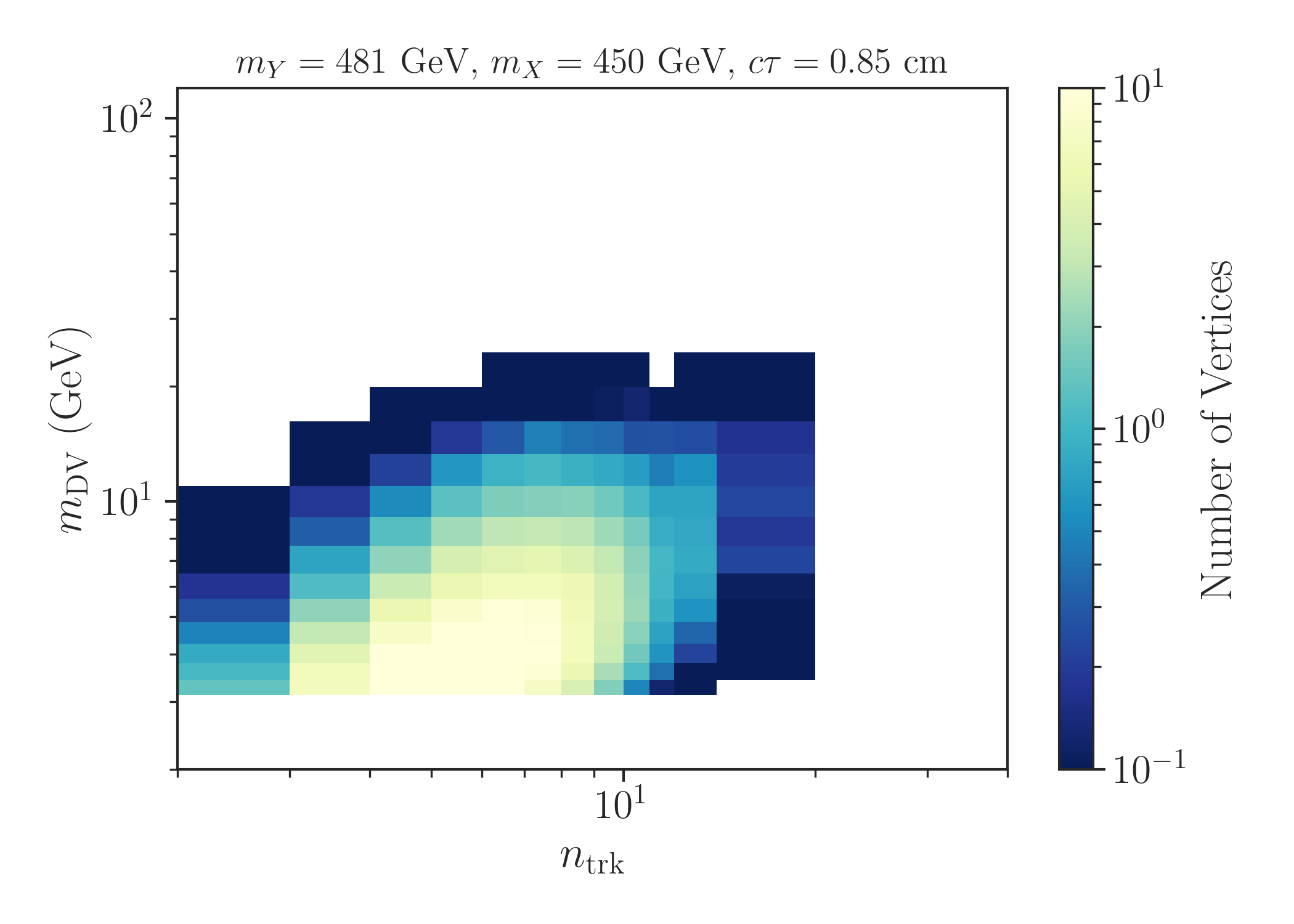}
    \caption{Distribution of $\mDV$ and $\nT$ for the benchmark $m_{Y} = 481$~GeV, $\Delta m = 31$~GeV, $\lambda_X = 3.9 \times 10^{-7}$.}
    \label{fig:mDVnTracks}
\end{figure}

Finally, searches involving displaced jets (or displaced vertices) signatures often require the displaced vertex (DV) to contain a minimum number of charged tracks ($\nT$) and a minimum invariant mass of the DV's tracks ($\mDV$) in order to reduce the physical and instrumental backgrounds. The corresponding distribution for the benchmark discussed above is shown in Fig.~\ref{fig:mDVnTracks}, where we see that most events have $\mDV \lesssim 5$~GeV and $\nT \simeq 5$. Once again the mass compression leads to a small $\mDV$, which makes it challenging to distinguish the signal from the SM background, \eg~SM bottom pair production. 

\section{Current LHC Constraints} \label{sec:currconstr}

To determine the region of parameter space still allowed by current LHC data, we consider the four searches listed in Table~\ref{tab:searches}: searches for heavy stable charged particles (HSCPs), disappearing tracks (DT), displaced jets (with and without missing energy), and prompt jets plus missing energy.
The searches have been recast using the information and parameterized efficiencies provided by the experimental collaborations\footnote{A validation of the ATLAS LLP searches can be found in Ref.~\cite{llpRepo}.}
and the event generation tools mentioned in Sec.~\ref{sec:lhc}. In addition, the recasting of the CMS disappearing track searches~\cite{CMS-EXO-19-010,CMS-EXO-16-044} was done using the \textsc{MadAnalysis\,5} SFS framework~\cite{Araz:2021akd} based on a previous implementation~\cite{DVN/P82DKS_2021} of this search available in the \textsc{MadAnalysis\,5} Public Analysis Database.
The relevant event selection criteria for all searches we discuss in the following can be found in Appendix~\ref{app:cfl}.

\begin{table}[h!]
  \centering
  \begin{tabular}{l|c|r}
ID   &  Signature  & Luminosity \\
\toprule
\href{https://atlas.web.cern.ch/Atlas/GROUPS/PHYSICS/PAPERS/SUSY-2018-42/}{ATLAS-SUSY-2018-42}~\cite{ATLAS-SUSY-2018-42} & HSCP & 139 fb$^{-1}$ \\ \hline
\href{https://cms-results.web.cern.ch/cms-results/public-results/publications/EXO-19-010/}{CMS-EXO-19-010}~\cite{CMS-EXO-19-010}   & \multirow{2}{*}{DT} & 101 fb$^{-1}$ \\
\href{https://cms-results.web.cern.ch/cms-results/public-results/publications/EXO-16-044/}{CMS-EXO-16-044}~\cite{CMS-EXO-16-044} & &  38 fb$^{-1}$ \\\hline
\href{https://atlas.web.cern.ch/Atlas/GROUPS/PHYSICS/PAPERS/SUSY-2016-08/}{ATLAS-SUSY-2016-08}~\cite{ATLAS-SUSY-2016-08} & DV  plus $E_\trans^\miss$ & 32.8 fb$^{-1}$ \\ \hline
\href{https://atlas.web.cern.ch/Atlas/GROUPS/PHYSICS/PAPERS/SUSY-2018-13/}{ATLAS-SUSY-2018-13}~\cite{ATLAS-SUSY-2018-13} & DV plus jets & 139 fb$^{-1}$ \\ \hline
\href{https://cms-results.web.cern.ch/cms-results/public-results/publications/EXO-20-004/}{CMS-EXO-20-004}~\cite{CMS-EXO-20-004}  & Jets plus $E_\trans^{\rm miss}$ &138 fb$^{-1}$ \\  \hline
\end{tabular}
\caption{Summary of the LHC Run II searches considered in this work. See text for details.}
\label{tab:searches}
\end{table}

\subsection{Heavy stable charged particles}

\begin{figure}
    \centering
    \includegraphics[width=0.48\textwidth,trim={0.45cm 0.6cm 0.1cm 0.4cm},clip]{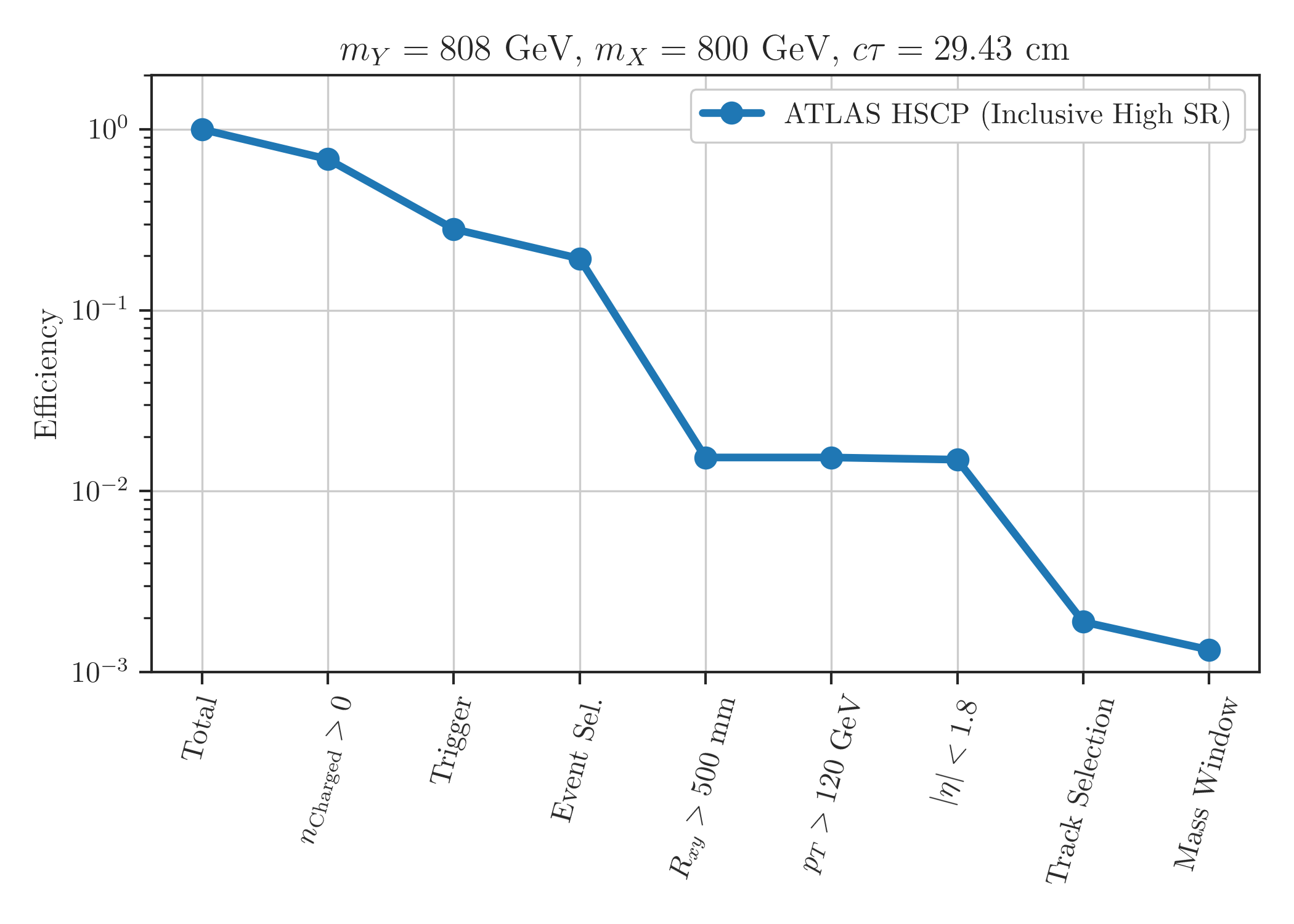}
    \caption{Cut flow for the ATLAS searches for heavy stable charged particles (HSCPs). The efficiency corresponds to the signal acceptance times efficiency after each selection cut. The results are for a benchmark model with  $m_{Y} = 808$~GeV, $\Delta m = 8$~GeV and $\tau(Y) = 0.98$~ns.}
    \label{fig:cutflowHSCP}
\end{figure}

Searches for HSCPs can be sensitive to the compressed region of the CDFO parameter space, where the $Y$ decay length takes its largest values (see Fig.~\ref{fig:cosmoparam}).
The ATLAS search from Ref.~\cite{ATLAS-SUSY-2018-42} looks for this type of signature and is able to constraint HSCPs with decay lengths $0.5$~m or higher. The trigger requires $E_{\trans,\calo}^\miss > 170$~GeV, where $E_{\trans,\calo}^\miss$ corresponds to the missing energy computed using the calorimeter deposits, which does not include the charged LLP if it decays outside the calorimeter. 
In Fig.~\ref{fig:cutflowHSCP}, we show the signal efficiency after the relevant cuts for a benchmark with small $\Delta m$ and $c \tau \simeq 0.3$~m.
We see that the requirement of charged R-hadrons only reduces the signal to about 70\%, while the trigger and event selection further reduces it to 20\%. 
Although the decay length is close to the largest ones occurring in the considered parameter space, the cut on the LLP displacement ($R_{xy} > 500$~mm) has a significant impact, reducing the efficiency down to 1.5\%. Finally, the track and the mass window requirements result in a final efficiency of about 0.2\%. 
This value quickly drops for larger $\Delta m$ values, \ie~smaller lifetimes.
In Fig.~\ref{fig:excCurves} we show the region in parameter space consistent with conversion-driven freeze-out (black curve) and the region excluded by the HSCP search (purple curve). 
As in Fig.~\ref{fig:cosmoparam}, the $Y$ lifetime (or $\lambda_X$ coupling) for each point in parameter space is fixed by requiring the correct DM relic abundance.
As we can see, the search can only constrain the region with $\Delta m \lesssim 15$~GeV.
We point out that the recasting information provided by ATLAS only allows us to recast the inclusive signal regions, which slightly underestimates the constraints. Also, for $c \tau < 0.3$~m, mass windows are only defined for $m_{\LLP} \geq 200$~GeV. This  explains the sudden drop in sensitivity in this region.

\subsection{Disappearing tracks}

\begin{figure}
    \centering
    \includegraphics[width=0.48\textwidth,trim={0.44cm 0.55cm 0.1cm 0.4cm},clip]{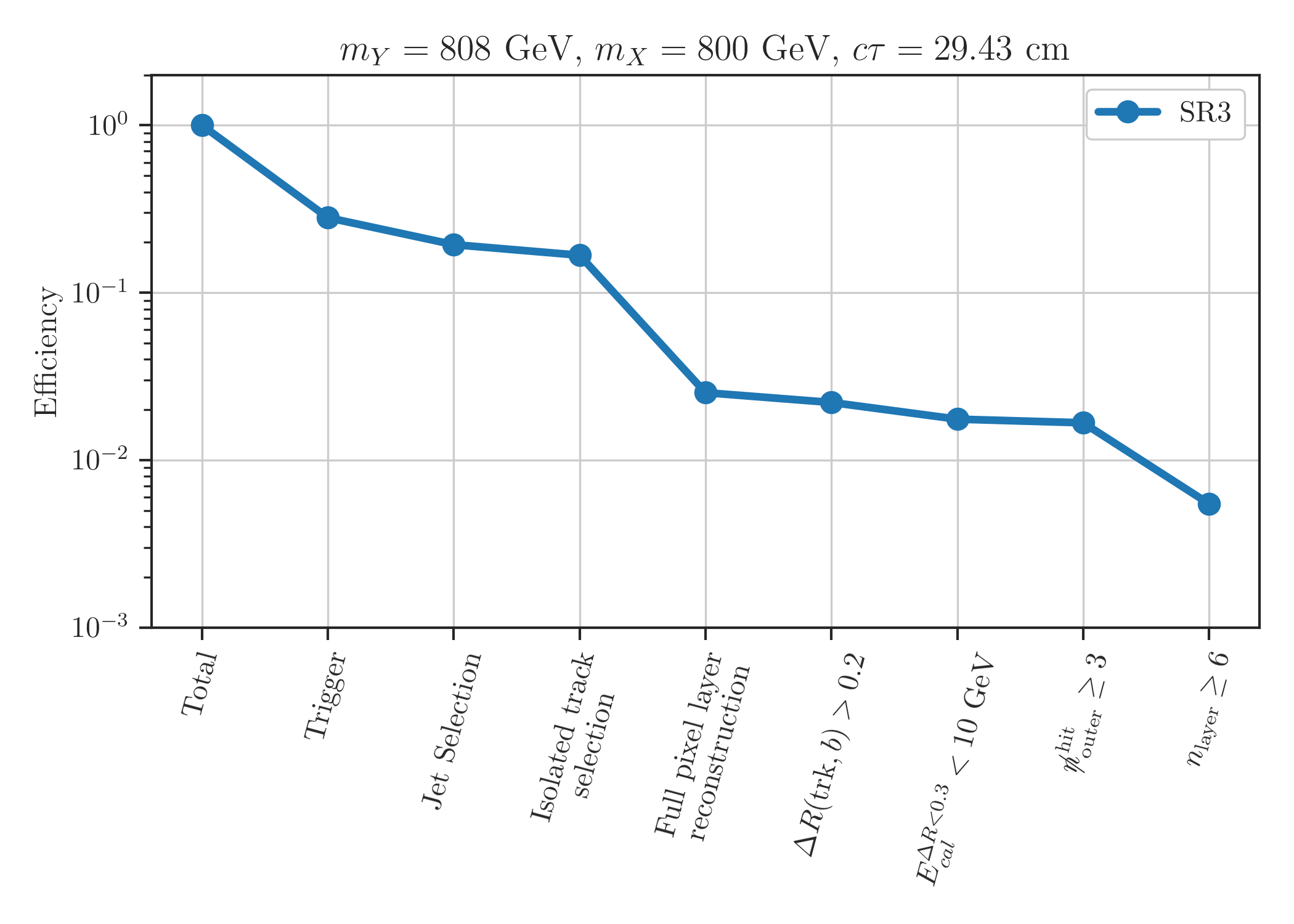}
    \caption{Cut flow for the CMS search for disappearing tracks (DT), using the same benchmark as Figure \ref{fig:cutflowHSCP} and the 2018B dataset. $\slashed{n}_\mathrm{outer}^\mathrm{hit}$ corresponds to the number of missing hits beyond the outermost recorded hit of a candidate disappearing track. $n_\mathrm{layer}$ quantifies the number of tracking layers with recorded hits.
    \label{fig:cutflowDT}}
\end{figure}

Disappearing track searches focus on charged particles with intermediate lifetimes ($c\tau \simeq 10 - 100$ cm), partially reconstructed in the tracker but 
without large calorimeter deposits.
The CMS disappearing track search~\cite{CMS-EXO-19-010} targets a model with a long-lived chargino decaying into a neutralino plus an ultra-soft pion. It spans three datasets: 2017, 2018A and 2018B (the latter two corresponding to pre and post-hadronic endcap calorimeter malfunction), with the recasting implemented in \cite{DVN/P82DKS_2021} also incorporating the 2015 and 2016 datasets~\cite{CMS-EXO-16-044}, thus including the full Run II luminosity.
In the conversion-driven freeze-out case, the compressed spectrum leads to an R-hadron decaying into DM plus a soft $b$-jet, which can meet the criteria for the search if 
the R-hadron is charged and the tracks associated with the $b$-jet have a large enough angular separation to the R-hadron.

The trigger for the DT search requires $p_\trans^\miss > 120$~GeV or $p_\trans^\miss > 105$~GeV and an isolated track with $p_\trans$ above 50 GeV.
Due to the possibility of identifying the candidate track as a muon, another trigger was implemented requiring $p_\trans^{\miss,\slashed{\mu}}>120$ GeV, where $\vec{p}_\trans^{\miss,\slashed{\mu}}$ is the negative vector sum over all non-muon transverse momenta in the event. 
Candidate tracks must be associated with the primary vertex and satisfy $p_\trans > 50$ GeV, $|\eta| < 2.1$.
In addition, the tracks must be isolated, \ie~the scalar $p_\trans$ sum of tracks inside a $\Delta R = 0.3$ cone of the candidate track has to be smaller than  5\% of the candidate $p_\trans$.\footnote{We point out that only tracks coming from the primary vertex are considered when computing the scalar sum. Therefore, displaced tracks, \eg~tracks associated with displaced $b$-jets do not contribute.}
Tracks are assigned a {\it missing hit} if they are reconstructed as passing through a functional tracker layer but with no associated recorded hit. Background mitigation requires candidate disappearing tracks to have hits in all Pixel layers and no missing hits between its inner and outermost recorded hits. A candidate track is tagged as {\it disappearing} if (1) the sum of all calorimetric energy within $\Delta R < 0.5$ of the track ($E_{\calo}^{\Delta R < 0.5}$) is less than 10 GeV and (2) it has at least 3 missing hits. 
Signal regions are then defined by the number of layers hit by the disappearing track: $n_\mathrm{layer}=4$ (SR1), $n_\mathrm{layer}=5$ (SR2) and $n_\mathrm{layer}\geq 6$ (SR3).

It is important to note that the standard Wino/Higgsino signal assumed in DT searches can differ significantly from the signal considered here.
While the chargino decays to DM and an ultra-soft pion, the $Y$ mediator decays to DM plus a $b$ quark, which can carry some visible energy depending on $\Delta m$.
Some of the tracks generated by the $b$-hadrons and their decays can have hits along the candidate track direction, thus reducing the number of missing hits. In order to take this effect into account, we impose an additional isolation cut: $\Delta R (\trk, b) > 0.2$, where $b$ is one of the $Y$ decay daughters. Although this is likely conservative, it allows us to apply the DT search to the conversion-driven freeze-out signal.

Following Ref.~\cite{CMS-EXO-19-010}, each signal region is treated as uncorrelated and combined across different datasets weighted by luminosity. From the resulting combination, the nominal exclusion limit is taken as the most stringent among the ones computed for each of the three regions. 
The SR3, corresponding to the longest decay lengths, has the most exclusion power due to the small expected background.

The impact of the DT selection cuts on the signal efficiency is shown in Fig.~\ref{fig:cutflowDT}. The trigger and jet selection reduces the signal to about 20\%. 
Although the isolated track selection is relevant for removing fake disappearing track candidates, it has no significant impact on the signal.
The most relevant cuts refer to the number of hits in the pixel and outer layers.
The full pixel layer reconstruction and the minimum number of missing hits typically imply that the DT search is sensitive to decay lengths (in the lab frame) between 16~cm and 100~cm.
For this particular benchmark, requiring that all the pixel layers have a hit reduces the efficiency to about 2\%, while the requirement for hits in at least 6 layers (for SR3) further reduces the efficiency to $\sim 0.5\%$.
We also see that the cut $\Delta R(\trk, b) >0.2$ imposed on the signal to ensure the disappearing condition has a small effect on the signal, rejecting only around 10\% of the events.
Comparing Figs.~\ref{fig:cutflowHSCP} and \ref{fig:cutflowDT}, we see that the signal efficiency for the DT search is almost an order of magnitude larger than the corresponding efficiency for the HSCP search.

The resulting 95\% CL exclusion limit is shown in Fig.~\ref{fig:excCurves} as the green shaded region.
For $\Delta m \simeq 6$~GeV, where the $Y$ proper decay length is larger than 30~cm, DM masses up to 1100 GeV are excluded. 
As $\Delta m$ increases and lifetimes become shorter, the signal regions SR1 and SR2, which target smaller tracks, become the most relevant. 
Since the exclusion power for these regions is smaller, the range of excluded DM masses quickly drops and, for $\Delta m > 25$~GeV, the DT search is no longer sensitive. We also see that, for all the parameter space considered, the DT track is more sensitive than the HSCP search.

\begin{figure*}[t]
    \centering
    \includegraphics[width=0.48\textwidth,trim={0.45cm 0.54cm 0.1cm 0.4cm},clip]{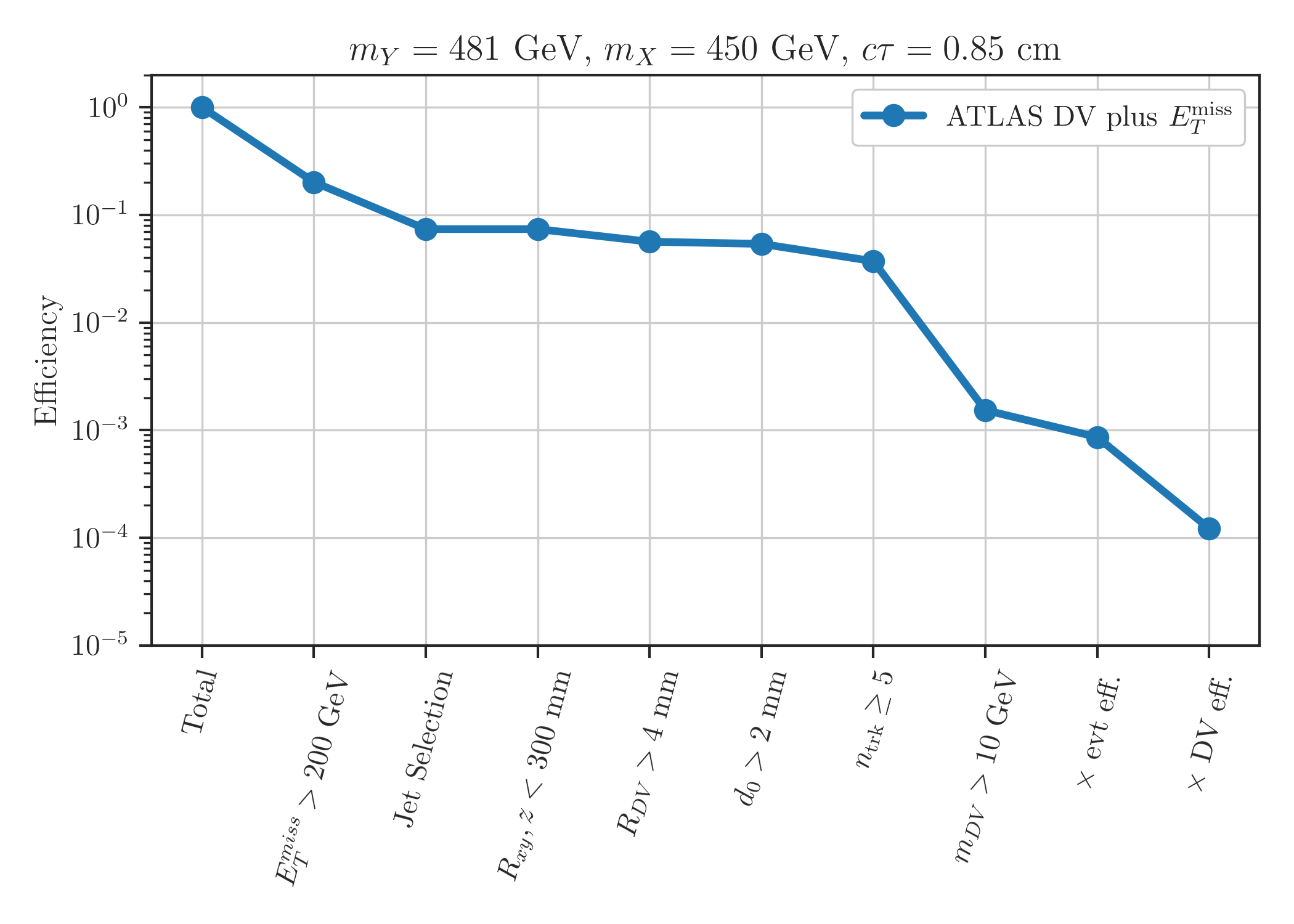} \hspace{3pt}
    \includegraphics[width=0.48\textwidth,trim={0.45cm 0.54cm 0.1cm 0.4cm},clip]{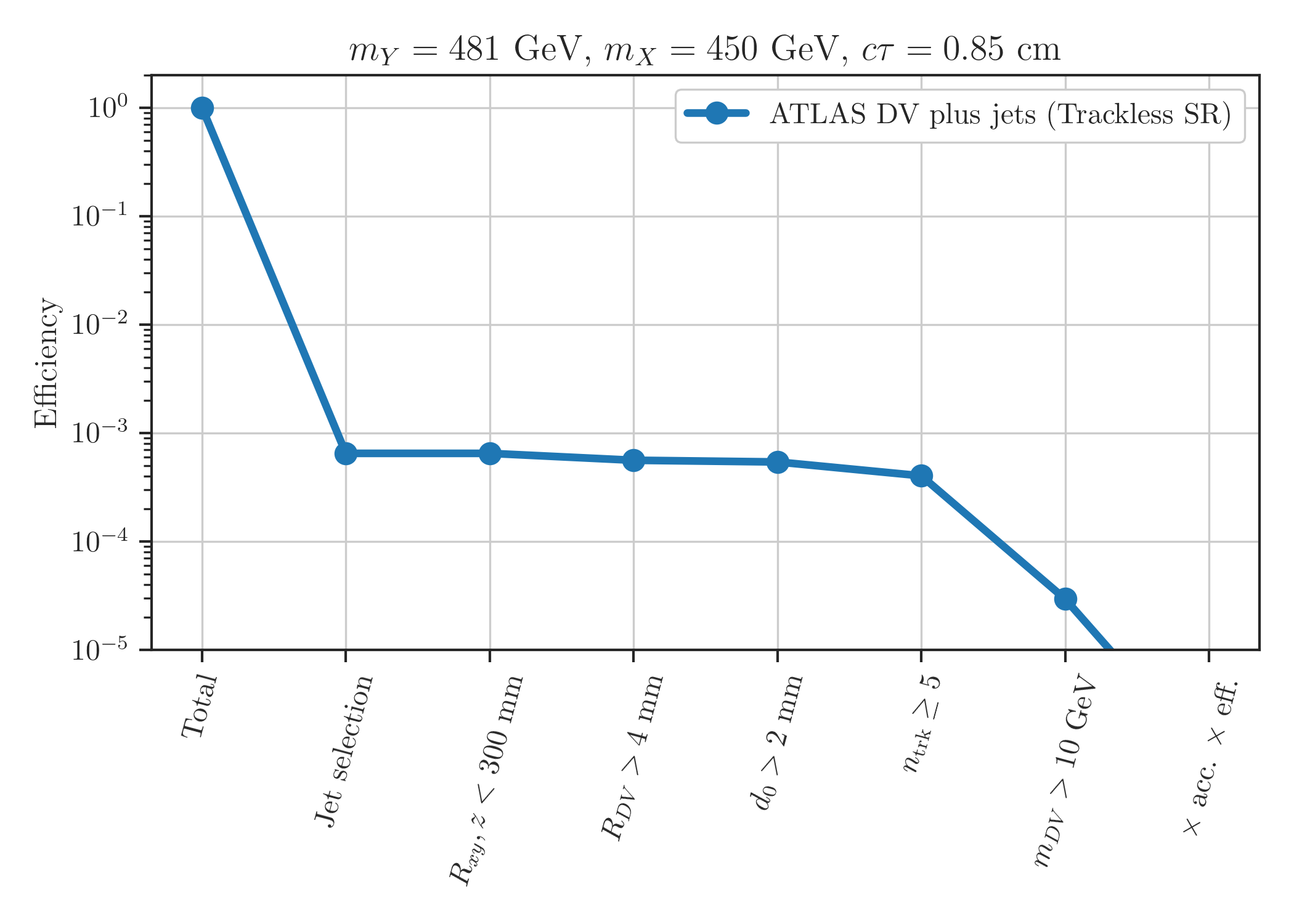}
    \caption{Cut flow for the ATLAS searches for displaced vertices plus $\met$ (left) and displaced jets (right). The efficiency corresponds to the signal acceptance times efficiency after each selection cut. The results are for a benchmark model with  $m_{Y} = 481$~GeV, $\Delta m = 31$~GeV and $\tau(Y) = 0.03$~ns.}
    \label{fig:cutflowATLAS}
\end{figure*}

\subsection{Displaced vertices}
\label{sec:currentDV}

The ATLAS displaced vertex plus $\met$ search~\cite{ATLAS-SUSY-2016-08} focuses on long-lived particles decaying to hard jets and invisible states. The main selection cuts include a hard missing energy cut, $\met > 200$~GeV, required by the trigger.
In addition, at least two jets with a small contribution from primary vertex (PV) tracks are required in the event.
The signal region is then defined by the DV selection, which includes the $\mDV > 10$~GeV and $\nT \geq 5$ cuts.

To understand how the CDFO signal efficiency (acceptance times efficiency) is affected by these cuts, we show the efficiency after the main analysis cuts in the left panel of Fig.~\ref{fig:cutflowATLAS}. As expected, the missing energy and jet selections (required by the trigger) are very stringent and already reduce the signal to 10\%. The next relevant cut is the requirement $\mDV > 10$~GeV, 
which reduces the signal to about 0.1\%. This drastic reduction is expected from the distribution in Fig.~\ref{fig:mDVnTracks}, which shows that most of the signal events have $\mDV < 5$~GeV.
After including the ATLAS efficiencies for the event selection and displaced vertex reconstruction the final signal efficiency is very small, $\epsilon_{\rm signal} = 10^{-4}$.
Such small efficiencies considerably restrict the region of the CDFO parameter space which can be probed by this analysis, as shown by the red curve in Fig.~\ref{fig:excCurves}. 
As we can see, only the region with high values of $\Delta m$ can be probed by the displaced jet search, since in this region it is easier to satisfy the $\mDV > 10$~GeV requirement.
The search loses some sensitivity at the highest $\Delta m$ values just at the edge of the parameter space, because in this region the $Y$ lifetime drops quickly, resulting in very small displacements.

As the recasting of the DV plus $\met$ search has only been validated within the split supersymmetry model considered in the original search and for larger mass splittings (see the validation material in Ref.~\cite{llpRepo}), the application to our scenario involves significant uncertainties. Even within the split supersymmetry model the  recasting shows deviations towards the region of small mass splittings overestimating the sensitivity up to about $\sim 50\%$ (for $\Delta m =  100\,$GeV). For illustration, we also show the exclusion curve assuming a 50\% reduction of the signal efficiency as the red dotted line in Fig.~\ref{fig:excCurves}. In this case the displaced vertex search can only exclude a very small strip of the parameter space.

We point out that the displaced jets plus $\met$ search discussed above does not make use of the full Run II luminosity. A search for DV using the full Run II data is available~\cite{ATLAS-SUSY-2018-13}, but targets events with hard displaced jets and imposes no $\met$ requirement. The main trigger for this search requires a large number of hard jets (see  Appendix~\ref{app:cfl}). In addition, displaced vertices are required to satisfy the same $\mDV$ and $\nT$ cuts as in the previous search. In the right panel of Fig.~\ref{fig:cutflowATLAS}, we show the signal efficiency for the DV plus jets search assuming the same benchmark model as before. In this case, since the signal only contains soft displaced jets, the jet selection already drastically reduces the signal efficiency to 0.01\%. Once the additional cuts are included, the signal drops to extremely small values and this search is not able to exclude any region of the CDFO parameter space.

\subsection{Jets plus missing energy}

\begin{figure}
    \centering
    \includegraphics[width=0.48\textwidth,trim={0.44cm 0.55cm 0.1cm 0.4cm},clip]{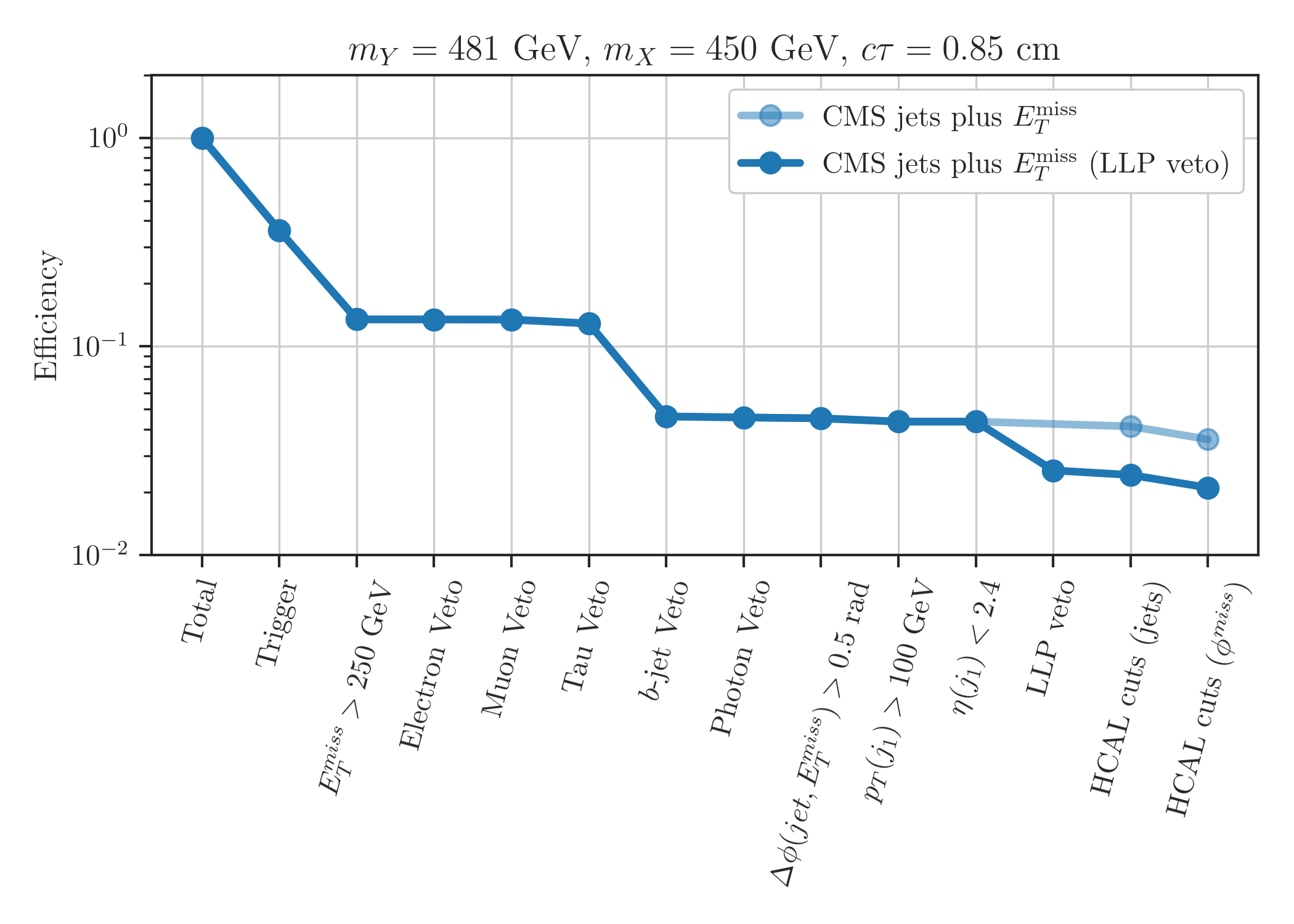}     
     \caption{Cut flow for the CMS jets plus $\met$ search with (dark blue) and without (bright blue) the LLP veto.}
    \label{fig:cutflowCMS}
\end{figure}

In the LLP searches discussed above, the requirement of large lifetimes, large DV invariant mass or multiple hard jets significantly reduces the signal efficiency in most of the CDFO parameter space. However, since the displaced jets are soft, we expect that searches for prompt (ISR) jets plus $\met$ can be sensitive to the CDFO signal as well.
Here we consider the CMS search for missing energy with at least one hard jet~\cite{CMS-EXO-20-004}.
The search is sufficiently inclusive, requiring only $\met > 250$~GeV and one jet with $p_\trans > 100$~GeV.
The search also includes a veto on $b$-jets with $p_\trans > 20$~GeV, but since the displaced $b$-jets present in the signal are usually soft, they can often satisfy this requirement.
When recasting this search for the signal considered here, one must be careful about how the presence of displaced jets can impact the signal efficiency.
If the displaced tracks are sufficiently soft and displaced, these can be considered as pile up and removed from the event. This removal, however, depends on the primary vertex reconstruction software used by CMS and cannot be easily reproduced outside the collaboration. 
In addition, events with highly ionizing tracks decaying within or after the calorimeter may impact the $\met$ reconstruction depending on whether the tracks are reconstructed as muons or not.
Therefore we consider two approaches when recasting the CMS search:
\begin{enumerate}[(i)]
    \item \label{ass1} consider displaced jets as prompt and highly ionizing tracks are considered as $\met$ or
    \item \label{ass2} veto events with displaced vertices satisfying $p_\trans(j) > 20$~GeV and $R > 2$~mm and/or with at least one charged R-hadron with $R > 1$~m, where $R$ is the transverse displacement. We refer to this set of cuts as {\it LLP veto}.
\end{enumerate}

\begin{figure}
    \centering
    \includegraphics[width=0.48\textwidth,trim={0.4cm 0.4cm 0.4cm 0.4cm},clip]{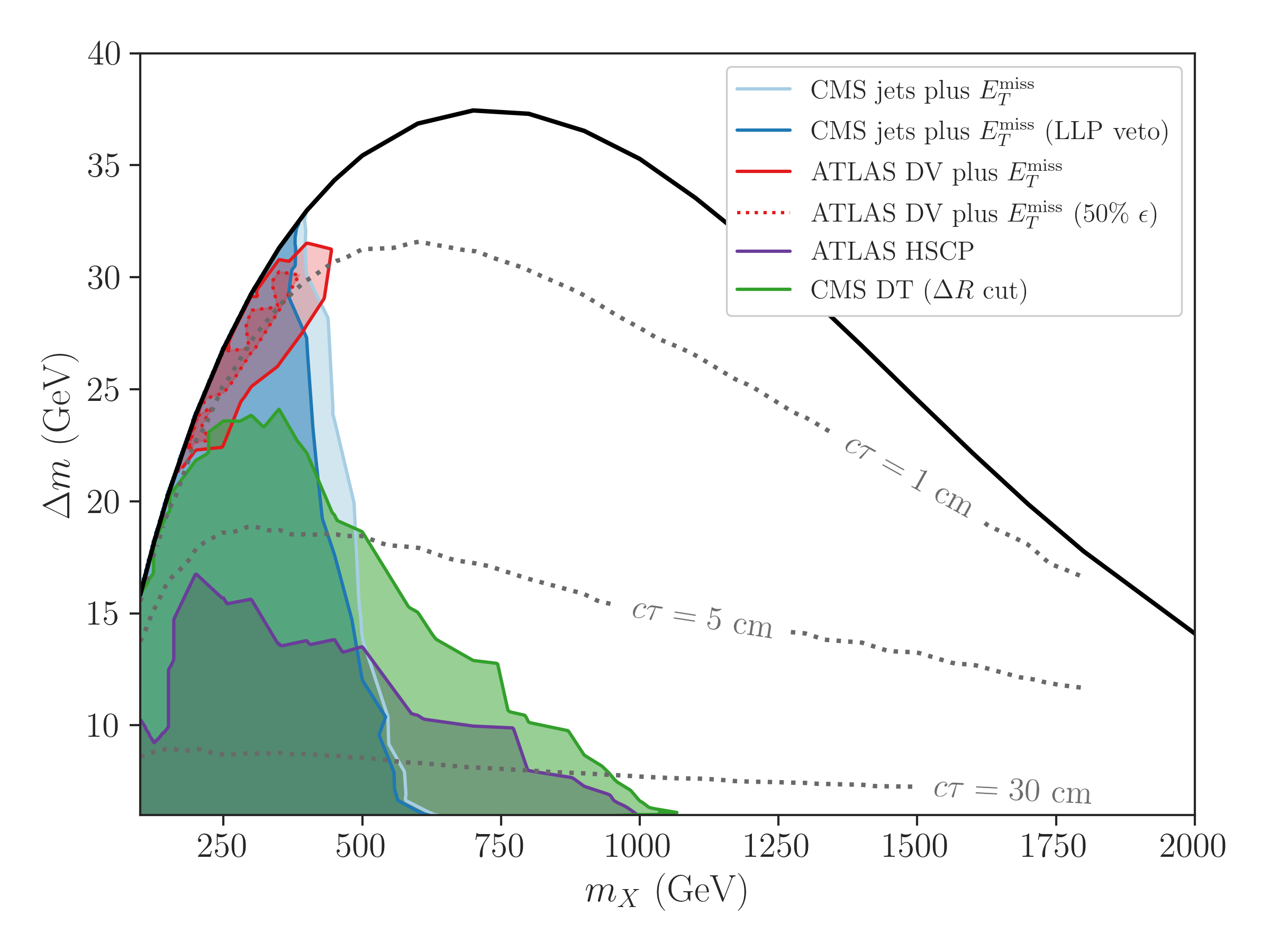}
    \caption{95\% C.L. exclusion curves for the LHC searches discussed in Sec.~\ref{sec:lhc}.}
    \label{fig:excCurves}
\end{figure}

The signal efficiencies for these two approaches are shown in Fig.~\ref{fig:cutflowCMS}, where we see that the trigger and $\met$ requirements reduce the signal to $\simeq 10$\%. Although the $b$-jet veto also has a significant impact on the efficiency, the final signal efficiency is $\sim 2$\%, which is considerably higher than for LLP searches.
We also see that the displaced jet veto imposed by assumption (\ref{ass2}) does not have a significant impact, reducing the final efficiency from 3\% to 2\%.
Despite the higher signal efficiency, the SM background in this case is sizeable, unlike the case of LLP searches.
The limits on the signal are computed using the CMS covariance matrix, which allows us to combine the 25 $\met$ bins considered by the search.
The results with and without the LLP veto are shown by the dark and light blue curves in Fig.~\ref{fig:excCurves}, respectively.
Although the CMS search does not exploit the displaced objects present in the signal, it still excludes a considerable fraction of the parameter space.
For $\Delta m \gtrsim 25$~GeV it is more sensitive than the LLP searches, except for a small region at $\Delta m \simeq 30$~GeV and $m_{X} \simeq 450$~GeV.

\section{Proposed Search for Soft Displaced Jets}
\label{sec:propse}

As discussed in Sec.~\ref{sec:currconstr}, the current LLP searches at the LHC are only marginally sensitive to the CDFO region with $\Delta m \gtrsim 25$~GeV.
In this region, the decay length is too small to provide sufficiently long tracks required by HSCP or DT searches. The displaced vertex searches, however, can be sensitive to decay lengths as small as a few mm.
Nonetheless, as discussed in Sec.~\ref{sec:currentDV}, the DV searches have small efficiencies due to two main factors: 
first, the stringent requirements on the jet activity and, secondly, the hard cut on the displaced vertex invariant mass, $\mDV$.
The first factor is only an issue for the DV plus jets search, while the $\mDV > 10$~GeV cut significantly reduces the signal efficiency by at least one order of magnitude for both searches.
Therefore in this section, we investigate how relaxing the $\mDV$ cut can potentially increase the LHC sensitivity to conversion-driven freeze-out models.

The $\mDV$ requirement is important to reduce the instrumental background since hadronic interactions in the detector can lead to a substantial number of SM events with small $\mDV$~\cite{ATLAS-SUSY-2016-08}. 
Nonetheless, a slightly weaker requirement, such as $\mDV > 5$~GeV, can significantly increase the sensitivity to the signal while still maintaining a small background.
For convenience, we define the number of expected background 
events or observed data for a set of $\mDV$ and $\nT$ cuts as:
\begin{equation}
  N_{\SM,\obs}(a,b)  \equiv  N_{\SM,\obs}(\mDV > a \mbox{ GeV }, \nT \geq b)  .
\end{equation}
Since a proper determination of the SM background  for small $\mDV$ can only be done within the experimental collaboration, here we use the ATLAS data from Ref.~\cite{ATLAS-SUSY-2016-08} for a rough estimate of the background.
To this end, we first assume that the background $\mDV$ distribution does not change rapidly with the number of tracks. In particular, we assume that:
\begin{equation}
    \frac{N_\SM(5, 5)}{N_\SM(10, 5)} \simeq \frac{N_\SM( 5, 3)}{N_\SM(10, 3)}.
\end{equation}
Second, we assume that the measured number of displaced vertices agrees with the SM prediction, 
 \begin{equation}
     \frac{N_\SM(5, 3)}{N_\SM(10, 3)} \simeq\frac{N_\obs( 5,  3)}{N_\obs( 10, 3)},
\end{equation}
and therefore:
\begin{equation}
    N_\SM(5, 5) \simeq \frac{N_\obs(5, 3)}{N_\obs( 10, 3)} N_\SM(10, 5) .\label{eq:bgExtrapolation}
\end{equation}
The fraction in Eq.~\eqref{eq:bgExtrapolation} can be obtained from the ATLAS publicly available data~\cite{ATLAS-SUSY-2016-08}:
\begin{equation}
     \frac{N_\obs(5, 3)}{N_\obs( 10, 3)}  = \frac{28}{13} \simeq 2.15,
\end{equation}
while the background for the ATLAS signal region is $N_\SM(10, 5) = 0.02 \pm 0.02$. Therefore we estimate:
\begin{equation}
     N_\SM( 5, 5) \simeq 0.04 \pm 0.04,
\end{equation}
where we have assumed the background error to be dominated by the systematical uncertainties, hence the same relative uncertainty was applied to our estimated background. We point out that despite being a rough estimate, our results are not strongly sensitive to the background prediction since it is extremely small and consistent with zero events observed in the proposed signal region $\mDV > 5$~GeV and $\nT \geq 5$.

A second issue related to relaxing the $\mDV$ cut concerns the DV reconstruction efficiencies that are needed for estimating the signal and were only provided for $\mDV > 10$~GeV. There is no clear way to determine these efficiencies for  smaller $\mDV$ values. Accordingly, we consider two different scenarios. In the first one, we assume the respective DV reconstruction efficiencies to be constant and equal to the one at $\mDV = 10$~GeV, \ie~following the nearest neighbor extrapolation:
\begin{equation}
    \epsilon_\mathrm{DV}\!\left(5 < \dfrac{\mDV}{\text{GeV}} < 10,\nT\right) \simeq \epsilon_\mathrm{DV}\!\left(\dfrac{\mDV}{\text{GeV}} = 10,\nT\right)
\end{equation}
This assumption is likely conservative since the ATLAS efficiencies depend on the SR definition and suppress events close to the SR border. Hence, looser cuts would probably increase the efficiencies for events falling along the $\mDV = 10$~GeV and $\nT = 5$ border.
Therefore, we also consider a second, more optimistic scenario, in which the vertex efficiency for all $\mDV$ and $\nT$ values is given by the average ATLAS efficiency:
\begin{equation}
    \epsilon_\mathrm{DV}(\mDV,\nT) \simeq \bar{\epsilon}_\mathrm{DV}
\end{equation}
where the average is taken over the ATLAS SR ($\mDV \geq 10$~GeV and $\nT \geq 5$).

\begin{figure}
    \centering
    \includegraphics[width=0.48\textwidth,trim={0.4cm 0.4cm 0.4cm 0.4cm},clip]{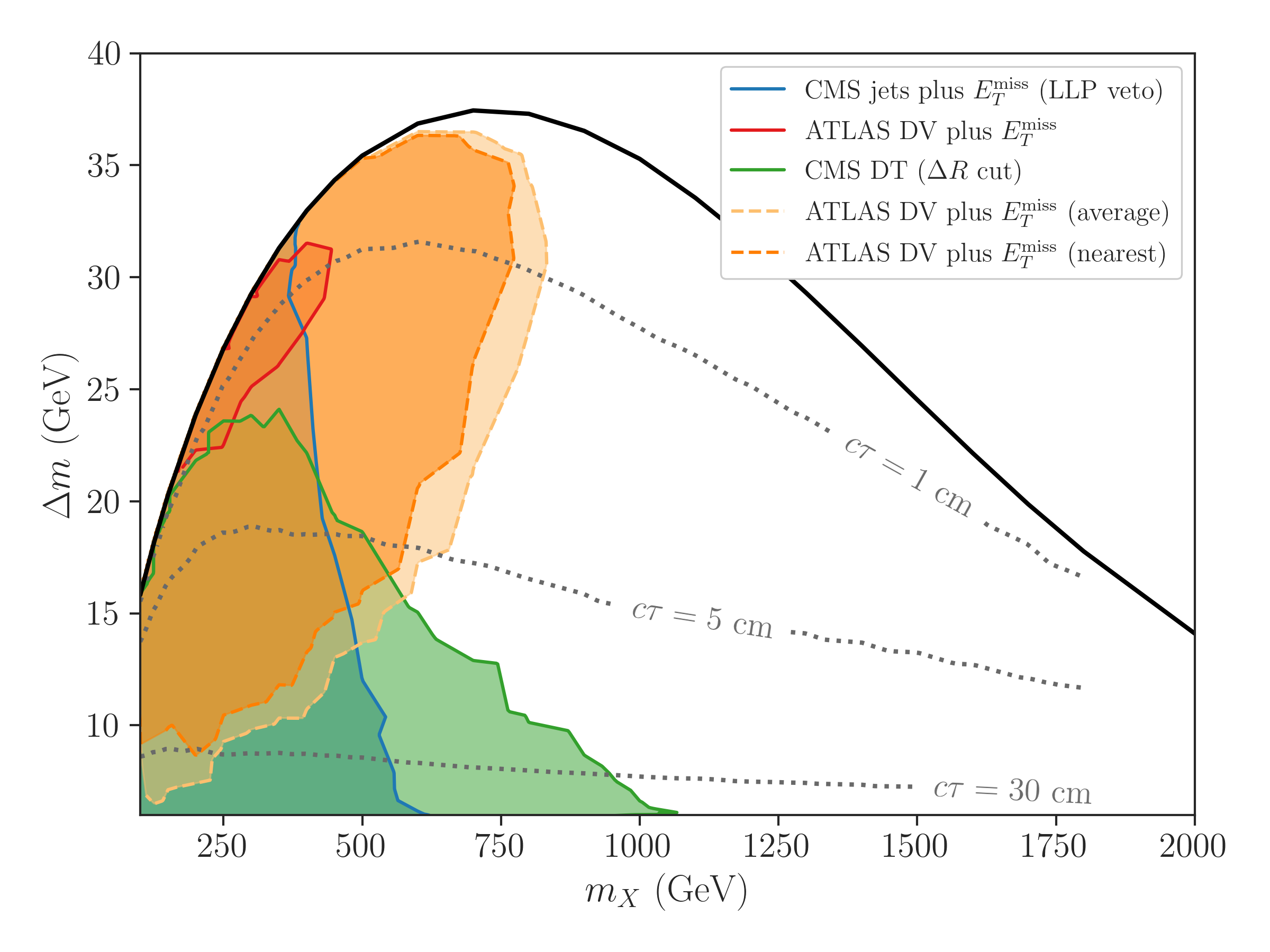}
    \caption{95\% C.L. exclusion curves for the LHC searches discussed in Sec.~\ref{sec:lhc}. The dashed curves for the ATLAS DV plus $\met$ search~\cite{ATLAS-SUSY-2016-08} correspond to the modified SR with $\mDV > 5$~GeV.}
    \label{fig:excCurves_mDV5}
\end{figure}

Under the above assumptions we can re-evaluate the reach of the DV plus MET search assuming a slightly smaller $\mDV$ cut. 
The resulting exclusion for the nearest neighbor extrapolation is shown as the dark orange curve in Fig.~\ref{fig:excCurves_mDV5}, while the bright orange curve shows the more optimistic estimate assuming the average efficiency.
Remarkably, the proposed signal region greatly enhances the exclusion at $\Delta m \gtrsim 25$~GeV excluding almost the entire parameter space with $m_X < 750$~GeV.\footnote{Note that within the assumptions of our analysis the significant increase in sensitivity is robust against the recasting uncertainties mentioned in Sec.~\ref{sec:currentDV}. While a 50\% reduction of the signal
efficiency reduces the excluded DM mass by around 50\,GeV, the relative gain due to the changed $\mDV$ cut remains roughly the same.}
Furthermore, all the LLP searches achieve a comparable reach on $m_X$, with the HSCP and DT searches probing the region with large lifetimes, while the DV search probes moderate lifetimes.

\section{Conclusion}
\label{sec:concl}

In this article, we studied the LHC signatures of a quark-philic $t$-channel mediator dark matter model in the conversion-driven freeze-out regime where very weak couplings explain the measured relic density. The scenario predicts long-lived particles with lifetimes ${\cal O}(\mathrm{mm})\lesssim c\tau \lesssim  {\cal O} (\mathrm{m})$ paired with a relatively small mass splitting between the long-lived particle and the dark matter particle it decays into. This combination has not been targeted by current LHC searches so far. 
Accordingly, we first investigated how current searches constrain the model. 
We considered searches for heavy stable charged particles, disappearing tracks, displaced jets (with and without missing energy) and prompt jets plus missing energy spanning the entire range from large to small decay lengths, respectively.

We found that the HSCP search, seeking the most inclusive signature of the above-mentioned ones can easily be reinterpreted within our model as it does not rely on the products or kinematics of the decay. However, while providing strong  constraints in the detector-stable limit, for the lifetime range focused on in this work, its 95\% CL exclusion limit is fully surpassed by limits from the considered disappearing track search. 
The latter constrains DM matter masses up to around 1100\,GeV for the smallest mass splitting considered in our analysis corresponding to a decay length of roughly a meter. 
The sensitivity of this search decreases towards larger mass splittings, \ie~smaller lifetimes. Still, it provides the strongest limit  up to a mass splitting of around 23\,GeV ($c\tau \sim$ a few cm) where searches for prompt jets plus missing energy take over. 
Since these prompt searches do not exploit the LLP nature, they provide a relatively weak limit, around $m_X\lesssim 400$\,GeV, despite the sizeable cross section for mediator pair production.

The region with decay lengths smaller than a few cm
would most promisingly be probed by searches for displaced vertices. However, current searches are designed to be most sensitive to hard displaced jets. 
Hence, we found that the search without $\met$ does not show any sensitivity to our scenario while the search including $\met$ can barely compete with the limit from prompt searches. This lack of sensitivity is due to the significantly softer jets arising from the long-lived particle decay in our scenario compared to the target scenario of the
searches. In particular, the cut on $\mDV>10\,$GeV removes most of our signal. 

We demonstrate that a significant improvement of sensitivity can be achieved by a slight reduction of this cut,  $\mDV>5\,$GeV.
Interestingly, according to our estimates based on the data reported by ATLAS, this reduction retains a sufficient suppression of the SM background.
Performing the analysis with the modified cut, we find that this analysis provides the strongest limits for the entire region above $\Delta m \sim 17\,$GeV, corresponding to $c\tau \simeq 10$~cm and reaches DM mass up to around $800\,$GeV in the regime of small lifetimes. These findings motivate including a signal region with loosened cuts in $\mDV$ in future analyses at the LHC\@.

The above discussion shows that exploiting the complementarity between different searches is key to fully probing the parameter space of conversion-driven freeze-out. Dedicated future searches may utilize the combination of the signatures of an anomalous (disappearing) track, a displaced vertex, and missing energy to further gain sensitivity in the intermediate range of lifetimes. 
Note that the combination of the former two shares similarities to a kinked track search.
Furthermore, timing information could be exploited as it is already done in searches for delayed jets~\cite{CMS:2019qjk}.
As a cosmologically viable realization of CDFO within the considered model restricts the new particles' masses to be below roughly $4\,$TeV, hence upcoming and future colliders provide promising prospects for fully probing the scenario with dedicated searches such as the one proposed here.

\vspace{48pt}

\section*{Acknowledgements}

We thank Mark Goodsell and Benjamin Fuks for helpful discussions.
J.H.~acknowledges support from the Alexander von Humboldt Foundation via the Feodor Lynen Research Fellowship for Experienced Researchers and Feodor Lynen Return Fellowship. A.L.~is supported by FAPESP grants no.~2018/25225-9 and 2021/01089-1\@.  L.M.D.R.~has received support from CNPq under grant no.~140343/2022-9, and from the European Union's Horizon 2020 research and innovation programme under the Marie Skłodowska-Curie grant agreement no.~860881-HIDDeN.

\begin{appendix}

\section{Cut flows}
\label{app:cfl}

In this appendix, we present the event selection cuts for the searches considered in this work. Note that only the most relevant cuts are shown
and that in some cases quality requirements for reconstructed objects are included as parametrized efficiencies provided by the collaborations.
Table~\ref{tab:atlasCutsHSCP} displays the HSCP selection from Ref.~\cite{ATLAS-SUSY-2018-42}, while 
Table~\ref{tab:cmsDT} shows the cuts for the CMS DT search from Ref.~\cite{CMS-EXO-19-010}.
The selection criteria for the displaced vertex searches from Refs.~\cite{ATLAS-SUSY-2016-08,ATLAS-SUSY-2018-13} are shown in Tables~\ref{tab:atlasCuts1} and \ref{tab:atlasCuts2}, respectively.
Finally, the CMS jets plus $\met$ selection is shown in Table~\ref{tab:cmsCuts}.

\vspace{17pt}

\begin{table}[h!]
\centering
\begin{tabular}{|l|c|c|}
\hline
\multicolumn{3}{|c|}{\large ATLAS HSCP}\\ \hline
 & Inclusive Low SR & Inclusive High SR \\ \hline
 \textbf{Trigger and} & \multicolumn{2}{|c|}{$E_{\trans,\calo}^\miss > 170$~GeV}  \\ 
 \textbf{Event Selection} & \multicolumn{2}{|c|}{}\\ \hline
 \multirow{3}{*}{\textbf{HSCP Selection}} & \multicolumn{2}{|c|}{$p_\trans > 120$~GeV} \\ & \multicolumn{2}{|c|}{$|\eta| < 1.8$} \\ & \multicolumn{2}{|c|}{$R_{xy} > 500$~mm} \\ \hline
\multirow{1}{*}{\textbf{Track Selection}} & $1.8 < \dEdx < 2.4$ & $\dEdx > 2.4$\\ \hline
\multirow{1}{*}{\textbf{Mass Window}} & $m \in [ m_{\rm Low}^{i},m_{\rm Low}^{i+1} ]$ &  $m \in [ m_{\rm High}^{i},m_{\rm High}^{i+1} ]$ \\ \hline
\end{tabular}
\caption{ATLAS selection for the HSCP search~\cite{ATLAS-SUSY-2018-42}. In the selection above $\dEdx$ is in units of MeV g$^{-1}$cm$^2$. The trigger and event selection, $\dEdx$ cuts and mass window selection were implemented using ATLAS parametrized efficiencies.\label{tab:atlasCutsHSCP}}
\end{table}

\begin{table}[h!]
\centering
\scalebox{1.0}{\begin{tabular}{|c|c|c|c|}
\hline
\multicolumn{4}{|c|}{\large CMS DT} \\ \hline
\multirow{3}{*}{\textbf{Trigger}} & \multicolumn{3}{|c|}{$p_\trans^\miss \geq 120$ GeV}\\ 
& \multicolumn{3}{|c|}{\textbf{or} $p_\trans^\miss \geq 105$ GeV, $n(\trk_{50}) \geq 1$}\\
& \multicolumn{3}{|c|}{\textbf{or} $p_\trans^{\miss,\slashed{\mu}} \geq 120$ GeV}\\ \hline
\multirow{3}{*}{\textbf{Jet Selection}} & \multicolumn{3}{|c|}{$n(j_{110}) \geq 1$, $|\eta(j_{110})| \leq 2.4$} \\
& \multicolumn{3}{|c|}{$\Delta\phi(\Vec{p}_\trans(j_1) ,\Vec{p_\trans}^\miss) > 0.5$} \\
& \multicolumn{3}{|c|}{$\Delta\phi(j_a,j_b) < 2.5$, $a\neq b$} \\\hline
\multirow{1}{*}{\textbf{Isolated Track}}& \multicolumn{3}{|c|}{$p_\trans > 55$ GeV, $|\eta| < 2.1$} \\ \multirow{1}{*}{\textbf{Selection}} 
& \multicolumn{3}{|c|}{$\left(\sum_{j\neq i} p_{\trans,j}\right)/p_{\trans,i} < 5\% $} \\ \hline
\textbf{Full Pixel Layer} & \multicolumn{3}{|c|}{\multirow{2}{*}{$\slashed{n}_{\rm pixel}^\mathrm{hit} = 0 $}} \\ \textbf{Reconstruction} & \multicolumn{3}{|c|}{}\\ \hline
\multirow{1}{*}{\textbf{No inner/middle}} & \multicolumn{3}{|c|}{\multirow{2}{*}{$\slashed{n}_{{\rm inner}}^\mathrm{hit} = 0$, $\slashed{n}_{{\rm mid}}^\mathrm{hit}=0$}}\\ \multirow{1}{*}{\textbf{missing hits}}& \multicolumn{3}{|c|}{}\\\hline
\multirow{2}{*}{\textbf{Displaced track}}
& \multicolumn{3}{|c|}{$|d_0(\trk)| <  0.2$ mm} \\
& \multicolumn{3}{|c|}{$|d_z(\trk)| <  5$ mm} \\ \hline
\textbf{Jet Veto} & \multicolumn{3}{|c|}{$\Delta R(\trk, j_{30}) > 0.5$} \\ \hline
\textbf{Lepton Veto} & \multicolumn{3}{|c|}{$\Delta R(\trk,l) > 0.15$} \\ \hline
\multirow{1}{*}{\textbf{Calorimeter}} & \multicolumn{3}{|c|}{\multirow{2}{*}{$E_{\calo}^{\Delta R < 0.3} < 10$ GeV}} \\ 
\multirow{1}{*}{\textbf{Isolation}} & \multicolumn{3}{|c|}{} \\ \hline
\textbf{Outer Missing Hits} & \multicolumn{3}{|c|}{$\slashed{n}_{\rm outer}^\mathrm{hit}\geq 3$}\\ \hline
\textbf{Signal Region 1} & \multicolumn{3}{|c|}{$n_{\rm layer} = 4$}\\ \hline
\textbf{Signal Region 2} & \multicolumn{3}{|c|}{$n_{\rm layer} = 5$}\\ \hline
\textbf{Signal Region 3} & \multicolumn{3}{|c|}{$n_{\rm layer}\geq 6$}\\ \hline
\end{tabular}
}
\caption{CMS selection for the disappearing tracks search~\cite{CMS-EXO-19-010}. Missing hits are split based on their relative position: between the primary vertex and the innermost recorded hit (inner), between the inner and outermost recorded hit (mid), beyond the outermost recorded hit (outer). \label{tab:cmsDT}}
\end{table}

\begin{table}[h!]
\centering
\begin{tabular}{|l|c|}
\hline
\multicolumn{2}{|c|}{\large ATLAS DV plus $E_\trans^{\rm miss}$}\\ \hline
\multirow{1}{*}{\textbf{$E_\trans^{\rm miss}$ Cut}} & $E_\trans^{\rm miss} > 200$ GeV\\ \hline
\multirow{3}{*}{\textbf{Jet Selection}} & $p_\trans(j_1) > 70$ GeV \\ & $p_\trans(j_2) > 25$~GeV \\ & $\sum_{\rm PV tracks} p_\trans < 5$ GeV  \\ & (only applied to 75\% of events)
\\ \hline
\multirow{5}{*}{\textbf{DV Selection}} & 4~mm $< R_\mathrm{DV} < 300$ mm \\ & $|z_\mathrm{DV}| < 300$ mm \\ & $\max(d_0(\mathrm{track})) >  2$ mm \\  & $n_{\rm tracks} > 5$ \\ & $\mDV > 10$~GeV
\\ \hline
\end{tabular}
\caption{ATLAS selection for the displaced jets plus missing energy search~\cite{ATLAS-SUSY-2016-08}. \label{tab:atlasCuts1}}
\end{table}

\begin{table}[h!]
\centering
\begin{tabular}{|l|c|c|}
\hline
\multicolumn{3}{|c|}{\large ATLAS DV plus jets}\\ \hline
 & High PT SR & Trackless SR \\ \hline
\multirow{4}{*}{\textbf{Jet Selection}} & $n(j_{250}) \geq 4$ & $n(j_{137}) \geq 4$ \\ 
& or $n(j_{195}) \geq 5$ & or $n(j_{101}) \geq 5$ \\
& or $n(j_{116}) \geq 6$ & or $n(j_{83}) \geq 6$ \\
& or $n(j_{90}) \geq 7$ & or $n(j_{55}) \geq 7$ \\ \hline
\multirow{2}{*}{\textbf{Disp. Jet Selection}} &  & $n(j_{70}) \geq 1$ \\
&  & or $n(j_{50}) \geq 2$ \\ \hline
\multirow{5}{*}{\textbf{DV Selection}} & \multicolumn{2}{|c|}{4~mm $ < R_\mathrm{DV} < 300$ mm} \\ & \multicolumn{2}{|c|}{$|z_\mathrm{DV}| < 300$ mm} \\ 
& \multicolumn{2}{|c|}{$\max(d_0(\mathrm{track})) >  2$ mm} \\
& \multicolumn{2}{|c|}{$n_{\rm tracks} > 5$} \\ 
& \multicolumn{2}{|c|}{$\mDV > 10$\,GeV} \\ 
\hline
\end{tabular}
\caption{ATLAS selection for the displaced jets plus multiple jets search~\cite{ATLAS-SUSY-2018-13}.\label{tab:atlasCuts2}}
\end{table}

\begin{table}[h!]
\centering
\begin{tabular}{|l|c|}
\hline
\multicolumn{2}{|c|}{\large CMS Jets plus $E_\trans^{\rm miss}$}\\ \hline
\multirow{1}{*}{\textbf{$E_\trans^{\rm miss}$ Cut}} & $E_\trans^{\rm miss} > 250$ GeV\\ \hline
\multirow{3}{*}{\textbf{Electron Veto}} & $n(e) = 0$ \\ & $p_\trans(e) > 10$~GeV \\ & $|\eta(e)| < 2.5$ \\ \hline
\multirow{3}{*}{\textbf{Muon Veto}} & $n(\mu) = 0$ \\ & $p_\trans(\mu) > 10$~GeV \\ & $|\eta(\mu)| < 2.4$ \\ \hline
\multirow{3}{*}{\textbf{Tau Jet Veto}} & $n(\tau) = 0$ \\ & $p_\trans(\tau) > 18$~GeV \\ & $|\eta(\tau)| < 2.3$ \\ \hline
\multirow{3}{*}{\textbf{B Jet Veto}} & $n(b) = 0$ \\ & $p_\trans(b) > 20$~GeV \\ & $|\eta(b)| < 2.4$ \\ \hline
\multirow{4}{*}{\textbf{Jet Selection}} & $\Delta \phi \left(j_1,E_\trans^{\rm miss}\right) > 0.5$ rad \\ & $p_\trans(j_1) > 100$~GeV \\ & $|\eta(j_1)| < 2.4$ \\ & Mitigation Cuts \\ \hline
\end{tabular}
\caption{CMS selection for the jets plus missing energy search~\cite{CMS-EXO-20-004}.\label{tab:cmsCuts}}
\end{table}

\section{Projection Estimate for the High Luminosity LHC}
\label{app:lhcHL}

The High Luminosity LHC (HL-LHC) aims for achieving a total luminosity $\mathcal{L} = 3$~ab$^{-1}$.
Since the LLP searches have a small background and their sensitivity is usually limited by the signal yields, the increase in luminosity can significantly increase their sensitivity to the CDFO scenario.

To estimate the projected HL-LHC reach we consider a luminosity of 3~ab$^{-1}$ and make the following simplifying assumptions:
\begin{itemize}
    \item the number of observed events agree with the expected SM background ($N_{\rm obs} = N_{\rm bg}$),
    \item the signal and background yields directly scale with luminosity as $N_{\rm bg,s} \propto \mathcal{L}$ and
    \item the background uncertainties are dominated by systematical errors and scale as $\delta_{\rm bg} \propto \mathcal{L}$. In addition, for the CMS jets plus $\met$ covariance matrix ($C_{ij}$), we assume $C_{ij} \propto \mathcal{L}^2$.
\end{itemize}

\begin{figure}
    \centering
    \includegraphics[width=0.48\textwidth,trim={0.4cm 0.4cm 0.4cm 0.4cm},clip]{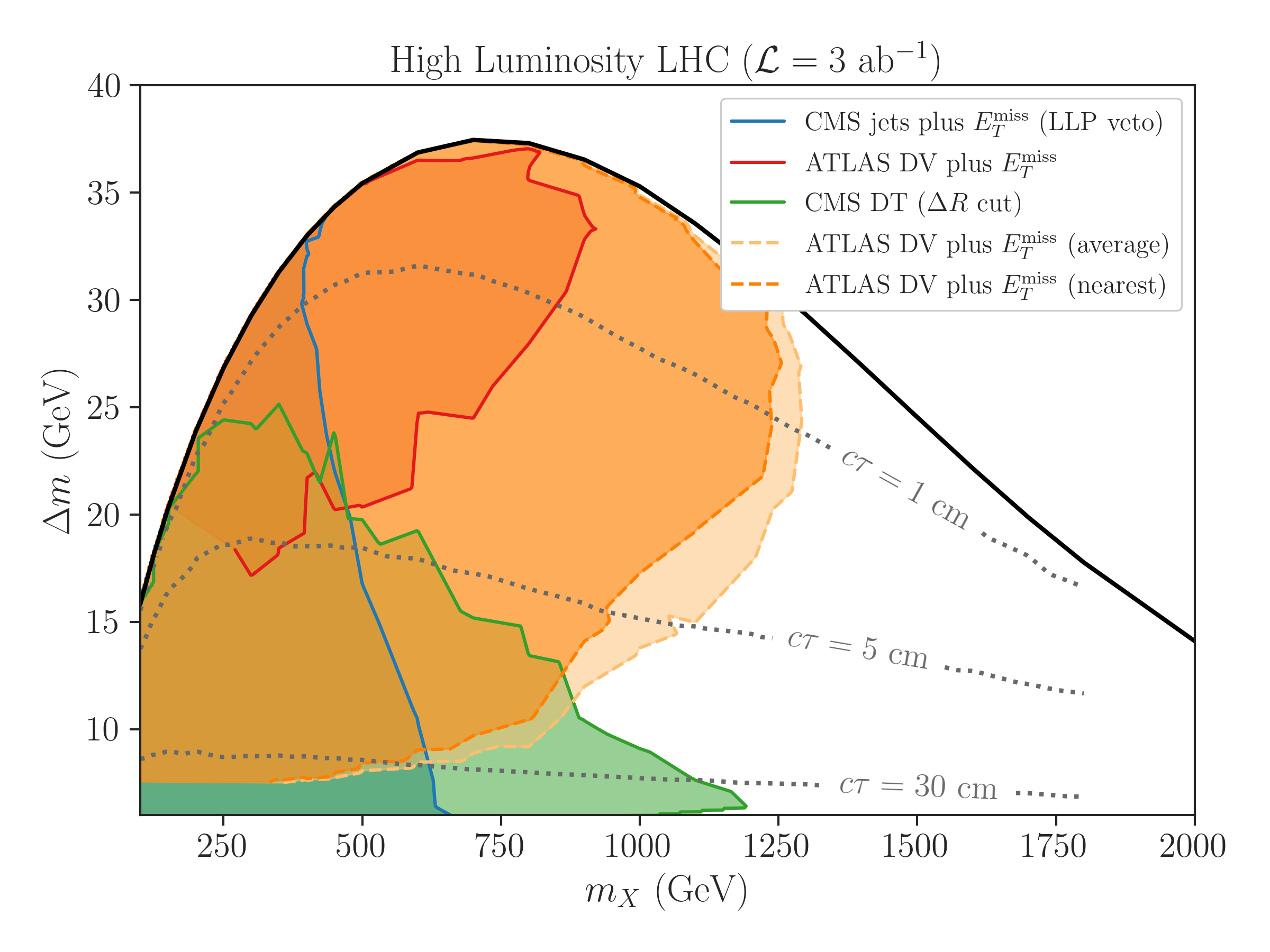}
    \caption{Estimated projections for the 95\% C.L.~exclusion limits at the High Luminosity LHC (see text for details).}
    \label{fig:excCurves_mDV5_HL}
\end{figure}

Note that detector upgrades and the high luminosity environment will likely affect the search sensitivities, but trying to account for these issues is beyond the scope of this work.
In Fig.~\ref{fig:excCurves_mDV5_HL}, we show the expected 95\% C.L.~exclusion limits obtained under the assumptions above. The HSCP reach is subdominant in all of the displayed parameter space, so we do not show it here.
Note that the estimated gain in sensitivity for the CMS jets plus $\met$ search is much smaller than the gain for the LLP searches.
This behavior is expected, since the LLP searches are limited by statistics, while the CMS search has large signal and background rates at current luminosities  already.
Finally, we point out that the estimated gain in the exclusion limit from a looser $\mDV$ cut has an equally significant impact at the HL-LHC, extending the overall reach to up to $m_X \simeq 1.2$~TeV.
It is therefore expected that a large portion of the allowed parameter space could be probed at the HL-LHC.

\end{appendix}

\newpage
\clearpage
\bibliographystyle{utcaps_mod}
\bibliography{bib}

\end{document}